\documentclass[aps,prd,amsmath,floats,floatfix,twocolumn,superscriptaddress,
  nofootinbib,showpacs]{revtex4-1} 

\usepackage{graphicx} 
\usepackage{bm}

\usepackage{hyperref}

\usepackage[usenames]{color}
\usepackage{ulem}
\newcommand{\Mirr}{M_{\text{irr}}}

\newcommand{\Sphere}{{\cal S}^3}
\newcommand{\Circle}{{\cal S}^2}
\newcommand{\PcutX}{{\cal P}_C}

\newcommand{\Caltech}{\affiliation{Theoretical Astrophysics 350-17,
    California Institute of Technology, Pasadena, CA 91125, USA}}
\newcommand{\CITA}{\affiliation{Canadian Institute for Theoretical Astrophysics,
        University~of~Toronto, Toronto, Ontario M5S 3H8, Canada}}

\begin{document}

\title{Simulations of unequal-mass binary black holes with spectral methods}

\author{Luisa T. Buchman}\Caltech
\author{Harald P. Pfeiffer}\CITA
\author{Mark A. Scheel}\Caltech
\author{B\' ela Szil\' agyi}\Caltech

\date{\today}

\begin{abstract}
  This paper presents techniques and results for simulations of
  unequal-mass, non-spinning binary black holes with pseudo-spectral
  methods.  Specifically, we develop an efficient root-finding
  procedure to ensure the black hole initial data have the desired
  masses and spins, we extend the dual coordinate frame method and
  eccentricity removal to asymmetric binaries. Furthermore, we
  describe techniques to simulate mergers of unequal-mass black
  holes. The second part of the paper presents numerical simulations
  of non-spinning binary black holes with mass ratios 2, 3, 4 and 6,
  covering between 15 and 22 orbits, merger and ringdown.  We discuss
  the accuracy of these simulations, the evolution of the (initially
  zero) black hole spins, and the remnant black hole properties.
\end{abstract}

% 04.25.D- Numerical relativity 
% 04.25.dg Numerical studies of black holes and black-hole binaries 
% 04.25.Nx Post-Newtonian approximation; perturbation theory; related approximations 
% 04.30.-w Gravitational waves (see also 04.80.Nn Gravitational wave detectors and experiments)
% 04.30.Db Wave generation and sources 
% 02.70.Hm Spectral methods 

\maketitle

%#########################
\section{Introduction}
%#########################

Numerical simulations of the inspiral and coalescence of two black
holes~\cite{Pretorius2005a} are an important tool for exploiting
upcoming gravitational wave detectors such as Advanced LIGO, VIRGO,
and LCGT/KAGRA~\cite{Barish:1999,Sigg:2008,
  Acernese:2008,Kuroda:2010,Somiya:2012}.  Increasingly larger sets of
simulations have begun to explore the parameter space of binary black
holes (BBHs),
%~\cite{pfeiffer-amaldi-talk}
most notably through the
NINJA~\cite{Aylott:2009tn,Aylott:2009ya,Ajith:2012tt} and
NRAR~\cite{NRARwebsite} collaborations.

One important subset of this parameter space comprises non-spinning
BBHs.  Head-on collisions have been studied
first~\cite{Anninos98,Moreschi99}, followed by simulations of inspiral
and coalescence of binaries that start in a quasi-circular orbit. One
well-studied phenomenon is the kick imparted to the remnant black hole
as a result of the collision of unequal-mass black
holes~\cite{Baker2006c,Gonzalez2007}; the form of this kick as a
function of the initial black hole masses is constrained by symmetry
considerations~\cite{Boyle2007b}.  Numerical simulations of
non-spinning BBH systems also formed the basis of analytic waveform
models and applications to gravitational wave data
analysis~\cite{Baker2008a,Buonanno2007,Ajith-Babak-Chen-etal:2007b,%
  Santamaria:2010yb,McWilliams2010b,HannamEtAl:2010,McWilliams2010a},
tuning of effective-one-body waveform
models~\cite{Damour2007a,DN2008,Buonanno:2009qa,PanEtAl:2011},
multipolar analysis~\cite{Berti-Cardoso-etal:2007,Schnittman2007}, and
investigations into the periastron advance of binary black
holes~\cite{Mroue2010,Tiec:2011bk}.  Recently, the range of mass
ratios covered by unequal-mass binaries has been extended to mass
ratios 10:1~ \cite{Gonzalez2008} and up to
100:1~\cite{Sperhake:2011ik,Nakano:2011pb,LoustoZlochower2010}.

Numerical simulations are still too computationally expensive to
include enough binary orbits for data analysis.  Therefore,
simulations are matched to post-Newtonian inspirals to obtain
``hybrid'' waveforms of sufficient length.  This matching must be done
early enough in the inspiral so that the post-Newtonian expressions
are still accurate.  During the last year, it has become increasingly
apparent that current numerical simulations are still not long enough
to provide an accurate match: the frequency range where post-Newtonian
and numerical waveforms are matched with each other is currently so
high that neglected higher-order terms in even state-of-the-art
post-Newtonian models lead to a noticeable impact on data
analysis~\cite{Hannam:2010,Santamaria:2010yb,Damour:2010,
  MacDonald:2011ne,OhmeEtAl:2011,Boyle:2011dy,Lovelace:2011nu}.

Unfortunately, the computational expense of a BBH inspiral is a steep
function of its initial frequency.  For instance, at lowest
post-Newtonian order~\cite{Blanchet2006}, a BBH inspiral starting at
an initial frequency $\Omega_i$ merges at a time
\begin{equation}
%T=5\eta^{3/5}(2\pi N)^{8/5} m,
T=\frac{5}{256}\, \eta^{-1} (M\Omega_i)^{-8/3}\,M,
\end{equation}
where $M$ is the total mass of the binary and $\eta$ its symmetric mass
ratio $\eta = M_1 M_2/(M_1+M_2)^2$. So even if the computational
expense were proportional to the evolution time $T$, it would be
expensive to significantly reduce $\Omega_i$; in practice the
situation is even worse because the computational expense (for a given
accuracy) increases superlinearly with $T$.  Therefore, long numerical
inspiral simulations (lasting $\gtrsim 10$ orbits) are rare, and are
generally available only for equal-mass binaries without
spin~\cite{Scheel2009}, or with equal spin magnitudes parallel to the
orbital angular momentum~\cite{Chu2009,Lovelace:2010ne}.

This paper revisits simulations of non-spinning unequal-mass binary
black holes, and describes accurate many-orbit waveforms,
including subdominant $(\ell,m)$ modes. 
Our simulations are performed with the Spectral
Einstein Code {\tt SpEC}~\cite{SpECwebsite}, a multi-domain
pseudo-spectral evolution code.  There are several motivations for
this work. First, we present an efficient technique to perform
10-dimensional root-finding that is necessary to construct BBH initial
data with specified masses and spins. Second, we present algorithms
for simulations of unequal-mass BBH systems with spectral methods.
Third, we present and carefully discuss a series of long
  duration, high-accuracy, unequal-mass non-spinning BBH simulations,
lasting between 15 and 22 orbits.  These simulations extend the
parameter space covered by spectral BBH evolutions, and improve in
length and accuracy already existing simulations which use alternative
numerical techniques.
The simulations presented here also provide additional
data points for remnant masses, spins and kick velocities, which we
compare with already published calculations and
analytical models.  Finally, we provide a study of tidal spin-up of
initially non-spinning black holes.

This paper is organized as follows.  Section~\ref{sec:Formalism}
presents details of our numerical implementation. First, the
quasi-circular, quasi-equilibrium initial
data~\cite{Caudill-etal:2006,Cook2004} require root-finding to adjust
free parameters so that after the initial data construction, the black
holes have specified masses and approximately zero spins -- we
introduce an efficient algorithm for performing this root-finding.
Second, we extend the dual-frame approach~\cite{Scheel2006} to
unequal-mass binaries, and discuss how we choose orbital parameters
that result in inspirals of orbital eccentricity $e<10^{-4}$.
Finally, we describe the handling of merger and ringdown, improving on
previous treatments~\cite{Scheel2009,Chu2009,Szilagyi:2009qz} of black
hole mergers performed with spectral multi-domain methods.
Section~\ref{sec:Results} presents numerical results for mass ratios
2, 3, 4, and 6.  These include results of convergence tests,
discussion of the black hole spin, detailed analysis of the leading
higher-order modes of the emitted gravitational waveform, and
discussion of the properties of the remnant black hole: mass, spin and
recoil velocity.  Section~\ref{sec:Discussion} summarizes and
discusses our main results.

We note that the simulations presented here have already been used in
the following published work: fitting
effective-one-body-models~\cite{PanEtAl:2011,Buonanno:2009qa} and
measuring the periastron advance for BBHs~\cite{Mroue2010}. They have
also been contributed to the Ninja2~\cite{Ajith:2012tt} and NRAR
projects~\cite{NRARwebsite}.  Further, the formalism for setting
initial data (cf. Sec.~\ref{sec:ID}) and for eccentricity removal
(cf. Sec.~\ref{sec:EccentricityRemoval}) was employed
in~\cite{Buonanno:2010yk,Tiec:2011bk}.

%%%%%%%%%%%%%%%%%%%%%%%%%%%%%%%%%%%%%%%%%%%%%%%%%%%%%%%%%%%%%%%%
\section{Formalism \& numerical methods}
\label{sec:Formalism}
%%%%%%%%%%%%%%%%%%%%%%%%%%%%%%%%%%%%%%%%%%%%%%%%%%%%%%%%%%%%%%%%

\subsection{Overview}

Our goal is to compute the last $\sim 20$ inspiral orbits, merger and
ringdown of binary black holes with mass ratio $q= M_1/M_2\ge 1$,
negligible spins of the black holes, and vanishingly small orbital
eccentricity.  This requires a rather complex sequence of steps:

\begin{enumerate} 
\item Choose the physical black hole masses $M_1$, $M_2$.
\item Decide on the initial coordinate separation $D_0$, and choose
  tentative values for the orbital frequency $\Omega_0$ and its
  time derivative, parameterized by $\dot{a}_0=\dot D(t)/D_0$ (for
  instance, based on post-Newtonian formulae).
\item \label{step:rootfinding} Fine-tune the 10 parameters that
    enter the initial data so that  the initial data
  contain black holes with desired masses, desired
  spins (here, zero), and vanishing center-of-mass motion.
\item \label{step:Evolution}Perform a short evolution lasting 2--3 orbits
   of the resulting initial-data set.
\item \label{step:EccRemoval} From the evolution in
    Step~\ref{step:Evolution}, extract information about the orbit of
  the binary and estimate the orbital eccentricity $e$.  If $e$ is
  unacceptably large, correct $\Omega_0$ and $\dot{a}_0$ and go back
  to step~\ref{step:rootfinding}. 
\item  If the orbital eccentricity $e$ is
  sufficiently small, continue the evolution through the remaining
  inspiral (for the current paper, we require $e<10^{-4}$).

\item \label{step:MergerRingdown} Simulate plunge, merger and ringdown.
\end{enumerate}
In order to accomplish our goal, we needed to make several refinements
to previous procedures used in {\tt SpEC} for
equal-mass~\cite{Pfeiffer-Brown-etal:2007,Boyle2007,Scheel2009,Chu2009}
and more generic (including $q=2$ unequal-mass)~\cite{Szilagyi:2009qz}
BBH simulations. These are: Step~\ref{step:rootfinding} was not
necessary in previous evolutions of simpler configurations, and is
explained in detail in Sec.~\ref{sec:ID} below.  Modifications to the
inspiral evolutions in Step~\ref{step:Evolution} are detailed in
Sec.~\ref{sec:EvolutionInspiralPhase}.  Eccentricity removal in
Step~\ref{step:EccRemoval} is generalized to mass ratios $q\neq 1$ in
Sec.~\ref{sec:EccentricityRemoval}.  Improvements to the merger and
ringdown phases (Step~\ref{step:MergerRingdown}) are described in
Sec.~\ref{sec:MergerRingdown}.  Finally, Sec.~\ref{step:CommonCode}
  summarizes code infrastructure that has not changed
  since earlier simulations; examples are apparent horizon finders and wave
  extraction.

%%%%%%%%%%%%%%%%%%%%%%%%%%%%%%%%%%%%%%%%%%%%%%%%%%%%%%%%%%%%%%%%
\subsection{Initial data}
\label{sec:ID}
%%%%%%%%%%%%%%%%%%%%%%%%%%%%%%%%%%%%%%%%%%%%%%%%%%%%%%%%%%%%%%%%

Quasi-equilibrium binary black hole initial
data~\cite{Cook2002,Cook2004,Caudill-etal:2006} are
constructed with the conformal thin sandwich
method~\cite{York1999,Pfeiffer2003b}.  This formalism results in a set
of five coupled non-linear elliptic equations, which are
solved numerically with a multi-domain
pseudo-spectral collocation method~\cite{Pfeiffer2003}.

As in earlier work, we employ the simplifying assumptions of conformal
flatness and maximal slicing.  Thirteen further real
parameters uniquely determine the complete initial data set. The
  orbital characteristics are determined by the three parameters $D_0$
  (coordinate separation), $\Omega_0$ (orbital frequency), and $\dot
  a_0$ (radial expansion factor);  their choice will be
discussed in detail in Sec.~\ref{sec:EccentricityRemoval}. The
remaining 10 parameters
\begin{equation}
\underline{u} = (r_1, r_2, \vec\Omega_1, \vec\Omega_2, X, Y)
\end{equation}
are the radii $r_1, r_2$ of the excision spheres, the angular
velocities of the horizons, $\vec \Omega_1$, $\vec \Omega_2$, and the
coordinate centers of the excision spheres, parameterized by $X$ and
$Y$ via $\vec c_1=(X, Y, 0)$ and $\vec c_2=(X-D_0, Y, 0)$.  We assume
that the black holes start in the $xy$ plane, with orbital angular
frequency parallel to the $z$-axis, i.e. the vectorial 
orbital frequency is written as $\vec\Omega_0=(0,0,\Omega_0)$.

The physical parameters (masses, spins, linear momentum) can only be
computed {\em after} the constraint equations are solved, whereas the
initial data parameters $\underline{u}$ must be chosen beforehand.
Therefore, 10-dimensional root-finding is required, to satisfy
\begin{align}
\underline{F}(\underline{u})\equiv&
(M_1 - M_1',\; M_2-M_2',\nonumber\\
& \;\,\vec{\chi}-\vec\chi_1',\; \vec{\chi} - \vec\chi_2',
P^x_{\rm ADM},\; P^y_{\rm ADM})
\nonumber\\
=&0.
\end{align}
Here, $M_{1,2}$, $\vec\chi_{1,2}$, and $\vec P_{\rm ADM}$ are,
respectively, the masses, dimensionless spins, and total linear
momentum, determined from the solution of the constraint equations,
whereas $M_{1,2}'$ and $\vec\chi_{1,2}'$ are the desired masses and
dimensionless spins of the black holes.  We also demand that the
initial ADM linear momentum $\vec P_{\rm ADM}$ vanish.  The $x$-- and
$y$-- components of $\vec{P}_{\rm ADM}$ are controlled by the choice of
$Y$ and $X$, respectively.  Its $z$--component $P^z_{\rm ADM}$ vanishes
by symmetry $z\to -z$ (in generic spinning cases, this will no longer
be the case).

In this paper, we will evolve only non-spinning black holes such that
$\vec\chi_{1,2}'=0$, but we present the root-finding for generic spins.

Each function evaluation $\underline{F}(\underline{u})$ requires
solving the elliptic constraint equations.  At high resolutions, this
requires a few hours of wall-clock time.  Because root-finding with
standard techniques such as the Newton-Raphson method~\cite{Press2007}
requires many function evaluations to compute the Jacobian, this would
result in inconveniently long run times\footnote{In earlier work on
  equal-mass binaries with equal aligned spins, this root-finding was
  not performed.  For those configurations, symmetry implies
  $r_1=r_2$, $\vec\Omega_1=\vec\Omega_2$ and $X=Y=0$.  The radii
  $r_1=r_2$ were chosen to be some fixed value, and the final black
  hole masses were simply measured (rather than controlled).  For the
  non-spinning simulation~\cite{Boyle2007}, $\vec\Omega_{1,2}$ were
  fixed at their values from quasi-circular non-spinning initial
  data~\cite{Caudill-etal:2006}; for the spinning
  simulation~\cite{Chu2009}, ~$\vec\Omega_1=\vec\Omega_2$ was chosen
  parallel to the $z$-axis, and the resulting black hole spin was just
  measured (rather than controlled).}.  To reduce computational
expense, we replace the exact Jacobian
$\partial{\underline{F}}/\partial\underline{u}$ by an {\em
  approximation} $\mathcal{J}_A$ and perform a Newton-Raphson
iteration employing $\mathcal{J}_A$.  That is, given parameters
$\underline{u}^{(k)}$, improved parameters are determined by
\begin{equation}\label{eq:NewtonRaphson}
  \Delta \underline{u}\equiv\underline{u}^{(k+1)} - \underline{u}^{(k)} =
  - \mathcal{J}_A^{-1}\underline{F}(\underline{u}^{(k)}),
\end{equation}
where $\mathcal{J}_A$ is evaluated at $\underline{u}^{(k)}$.

Efficiency of this technique hinges crucially on the quality of the
approximated Jacobian $\mathcal{J}_A$.  We compute $\mathcal{J}_A$
based on considerations that are valid for single black hole initial
data, and/or Newtonian gravity.  Specifically, for conformally flat
single black hole initial data with maximal slicing, the mass is
proportional to the radius of the excision sphere; therefore, we take
\begin{subequations}
\begin{equation}
\label{eq:JA_first}
\frac{\partial M_A}{\partial r_A}=\frac{M_A}{r_A}, \qquad A=1,2.
\end{equation}
Furthermore, for Kerr black holes with small spin, the dimensionless
spin parameter $\vec\chi$ is related to the angular frequency of the
horizon $\vec\Omega_H$ by $\vec\chi=4M\vec\Omega_H$, where $M$ is the
mass of the Kerr black hole.  For BBHs, the horizon frequency
$\vec\Omega_A$ measures spin in {\em addition} to co-rotation, so that
$\vec\chi_A=4M_A(\vec\Omega_A-\vec\Omega_0)$, from which follows
\begin{equation}
\frac{\partial\vec\chi_A}{\partial r_A}=\frac{\vec{\chi}_A}{r_A},
\qquad
\frac{\partial\vec\chi_A}{\partial\vec\Omega_A}=4M_A,\qquad A=1,2.
\end{equation}
Finally, in Newtonian gravity, the linear momentum is given by $\vec
P=M_1\vec\Omega_0\times\vec c_1 + M_2\vec\Omega_0\times\vec c_2$.
Substituting in $\vec\Omega_0=(0,0,\Omega_0)$, $\vec c_1=(X, Y, 0)$,
$\vec c_2=(X-D_0, Y, 0)$, one finds
\begin{align}
\frac{\partial P^x}{\partial r_1}&=-\frac{M_1}{r_1}\Omega_0 Y,\qquad
\frac{\partial P^x}{\partial r_2}=-\frac{M_2}{r_2}\Omega_0 Y,\\
\frac{\partial P^x}{\partial Y} &= -(M_1+M_2)\Omega_0,\\
\frac{\partial P^y}{\partial r_1}&=\frac{M_1}{r_1}\Omega_0 X,\qquad
\frac{\partial P^y}{\partial r_2}=\frac{M_2}{r_2}\Omega_0 (X-D_0),\\
\frac{\partial P^y}{\partial X} &= (M_1+M_2)\Omega_0.
\label{eq:JA_last}
\end{align}
\end{subequations}
Equations~(\ref{eq:JA_first})--(\ref{eq:JA_last}) are the only
non-zero components of $\mathcal{J}_A$.  Because the Jacobian is so
sparse, it is trivial to solve Eq.~(\ref{eq:NewtonRaphson}), and one
obtains:
\begin{subequations}
\begin{align}
  \Delta r_A &= - r_A \frac{M_A-M_A'}{M_A},\qquad\qquad\qquad\;\, A=1,2\\
  \Delta\vec\Omega_A &= -\frac{\vec\chi_A-\vec\chi_A'}{4M_A}+\frac{M_A-M_A'}{4M_A^2}\vec\chi_A,\quad A=1,2\\
\Delta X&= \frac{-P^y_{\rm ADM}}{(M_1+M_2)\Omega_0}\nonumber\\
&\quad +\frac{X (M_1\!-\!M_1')+(X\!-\!D_0)(M_2\!-\!M_2')}{M_1+M_2}\\
\Delta Y&= \frac{P^x_{\rm ADM}}{(M_1+M_2)\Omega_0}\nonumber\\
&\quad+\frac{Y (M_1-M_1'+M_2-M_2')}{M_1+M_2}
\end{align}
\end{subequations}
In these equations, primed quantities are the desired values, whereas
un-primed quantities are determined from the initial data computed
from parameters $\underline{u}^{(k)}$.

\begin{figure}
\centerline{\includegraphics[scale=0.5]{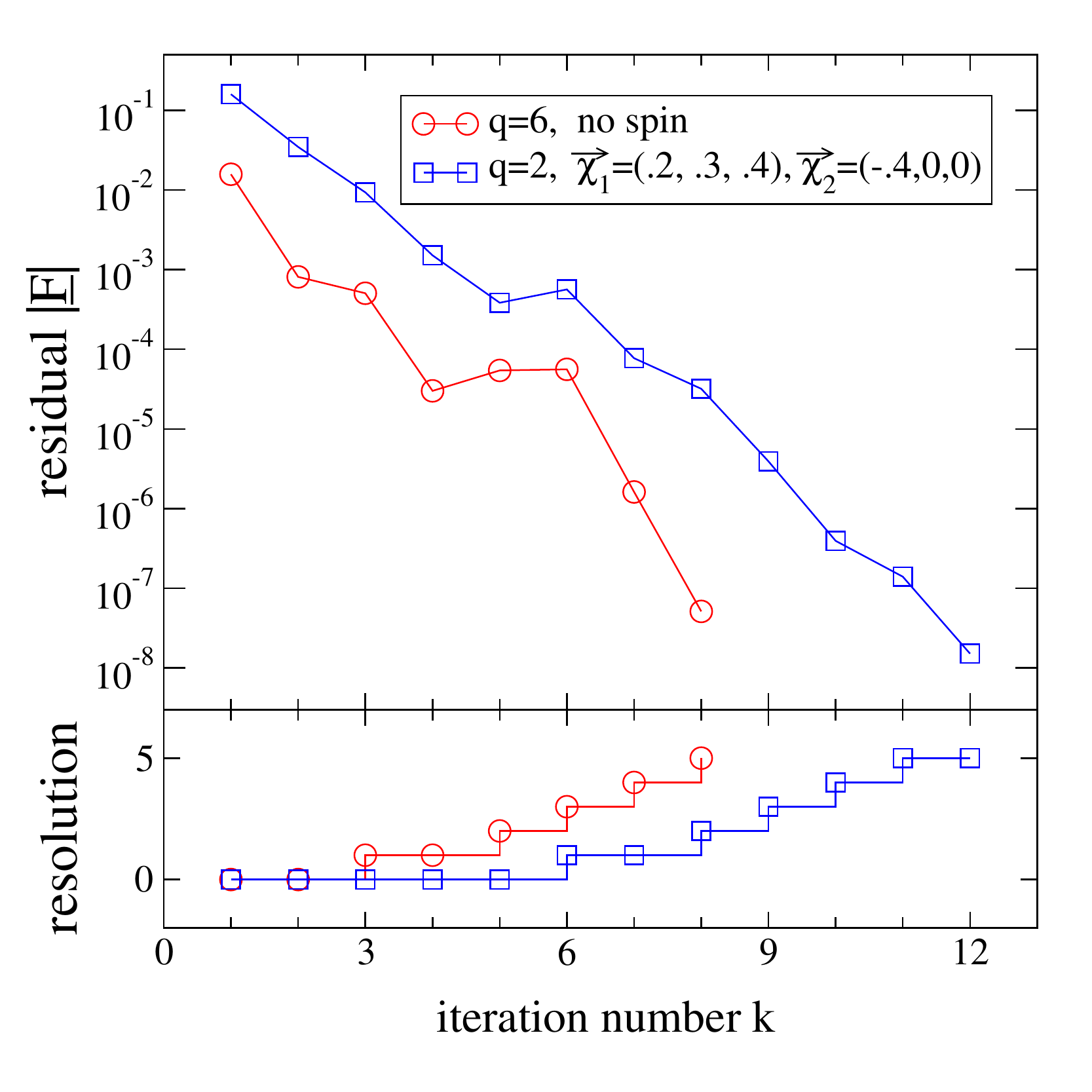}}
\caption{\label{fig:BBH_ID_Convergence} Residual $|\underline{F}|$ of
  the initial-data root-finding procedure (top panel) vs. iteration
  number.  The lower panel indicates the numerical resolution of each
  elliptic constraint solve, with 5 being highest resolution (each
  increase of this integer corresponds to adding a certain number of
  basis-functions, cf. Fig. 6 of~\cite{Cook2004}). }
\end{figure}

Fig.~\ref{fig:BBH_ID_Convergence} demonstrates the efficiency of this
procedure for two configurations.  During the first iterations of
root-finding, we solve the constraint equations only to lowest
resolution.  We begin to increase the resolution $k_{\rm Ell}$ of the
elliptic solver when the residual $|\underline{F}|$ falls within a
factor of $10^4$ of our target tolerance $10^{-7}$.  Because solving
the constraint equations at low resolution is very quick, the overall
cost of the root-finding is dominated entirely by the solutions of the
constraint equations at highest resolution, and thus, the entire
root-finding adds only a small amount of wall-clock time.

As is apparent in Fig.~\ref{fig:BBH_ID_Convergence}, the quadratic
convergence of Newton-Raphson algorithm is lost because of the
approximations entering ${\mathcal J}_A$.  We find roughly linear
convergence where each iteration reduces the error by a certain
factor.  The convergence rate depends on how closely $\mathcal{J}_A$
resembles the exact Jacobian.  Convergence is not exactly linear,
because we delay increasing the resolution of the elliptic solver
until as high $k$ as possible, to gain maximum speed-up from the lower
resolution solutions.

\subsection{Evolution of inspiral phase}
\label{sec:EvolutionInspiralPhase}

The Einstein evolution equations are solved with the pseudo-spectral
evolution code ({\tt SpEC}) described in Ref.~\cite{Scheel2009}.
This code evolves a first-order representation~\cite{Lindblom2006} of
the generalized harmonic
system~\cite{Friedrich1985,Garfinkle2002,Pretorius2005c} and includes
terms that damp away small constraint
violations~\cite{Gundlach2005,Pretorius2005c,Lindblom2006}.  The
computational domain extends from excision boundaries located just
inside each apparent horizon to some large radius, and is divided into
subdomains with simple shapes (e.g. spherical shells, cubes,
cylinders).  No boundary conditions are needed or imposed at the
excision boundaries, because all characteristic fields of the system
are outgoing (into the black hole) there.  The boundary conditions on
the outer boundary~\cite{Lindblom2006,Rinne2006,Rinne2007} are
designed to prevent the influx of unphysical constraint
violations~\cite{Stewart1998,FriedrichNagy1999,Bardeen2002,Szilagyi2002,%
  Calabrese2003,Szilagyi2003,Kidder2005} and undesired incoming
gravitational radiation~\cite{Buchman2006,Buchman2007}, while allowing
the outgoing gravitational radiation to pass freely through the
boundary. Interdomain boundary conditions are enforced with a penalty
method~\cite{Gottlieb2001,Hesthaven2000}.

The gauge freedom in the generalized harmonic formulation of
Einstein's equations is fixed via a freely specifiable gauge source
function $H_a$ that satisfies the constraint
\begin{equation}
  \label{e:ghconstr}
  0 = \mathcal{C}_a \equiv \Gamma_{ab}{}^b + H_a,
\end{equation}
where $\Gamma^{a}{}_{bc}$ are the spacetime Christoffel symbols.
During the inspiral, we choose $H_a$ as in
Refs.~\cite{Boyle2007,Scheel2009,Chu2009}.

In order to treat moving holes using a fixed grid, we employ multiple
coordinate frames~\cite{Scheel2006}: the equations are solved in an
`inertial frame' that is asymptotically Minkowski, but the grid is
fixed in a `grid frame' in which the black holes do not move.  The
motion of the holes is accounted for by dynamically adjusting the
coordinate mapping between the two frames\footnote{All coordinate
  quantities (e.g. trajectories, waveform extraction radii) in this
  paper are given with respect to the inertial frame unless noted
  otherwise.}.  This coordinate mapping differs from our earlier work,
and is described below in Sec.~\ref{sec:CoordMap}.

Furthermore, the choice of constraint damping parameters is important
for stability, and it discussed in Sec.~\ref{sec:ConstraintDamping}.

%%%%%%%%%%%%%%%%%%%%%%%%%%%%%%%%%%%%%%%%%%%%%%%%%%%%%%%%%%%%%%%%
\subsubsection{Dual-frames \& Control system}
\label{sec:CoordMap}
%%%%%%%%%%%%%%%%%%%%%%%%%%%%%%%%%%%%%%%%%%%%%%%%%%%%%%%%%%%%%%%%

{\tt SpEC} utilizes two coordinate
systems~\cite{Scheel2006}: grid coordinates $x^i$, in
which the domain decomposition is fixed, and inertial coordinates
$x^{\bar\imath}$, in which the black holes orbit around each other.
The mapping between these coordinate systems is chosen such that in
grid coordinates, the black holes remain centered on the excision
spheres.  In earlier simulations of equal-mass
binaries~\cite{Scheel2006,Pfeiffer-Brown-etal:2007,Boyle2007}, this
map was chosen to be a rotation and an overall scaling. Unequal-mass
binaries will acquire a kick in the orbital plane; therefore, we add a
translation to the mapping between inertial and grid coordinates:
\begin{equation}\label{eq:SpatialCoordMap}
x^{\bar\imath}=a(t) R^{\bar\imath}{}_i x^i + T^{\bar\imath}.
\end{equation}
Here, $a(t)$ is the overall scale factor,
$T^{\bar\imath}=(T^{\bar{x}},T^{\bar{y}},0)$ represents the
translation, and
\begin{equation}
R^{\bar\imath}{}_i=
 \left(\begin{aligned}
\mathbf{R}_\phi && \mathbf{0} \\
\mathbf{0\;\,} && 1
\end{aligned}
\right),
\qquad \mathbf{R}_\phi=
\left(\begin{aligned}
\cos\phi && -\sin\phi \\
\sin\phi && \cos\phi  
\end{aligned}
\right)
\end{equation}
is the rotation matrix for a rotation by the angle $\phi(t)$ about the
$z$-axis.  The rotation and translation act only on the $x-$ and
$y-$coordinates, because a non-spinning unequal-mass binary is, by
symmetry, confined to remain in the $xy$-plane\footnote{Spinning,
  unequal-mass binaries with both black hole spins
  {\em parallel} to the orbital angular momentum will also remain in a
  fixed orbital plane.  Our discussion applies equally well to these
  systems.}.

The mapping Eq.~(\ref{eq:SpatialCoordMap}) is determined by four
free-functions, $\lambda_\alpha\equiv\{a(t), \phi(t), T^{\bar x}(t),
T^{\bar y}(t)\}$ (where $\alpha$ labels the four functions).  The
functions $\lambda_\alpha(t)$ must be chosen dynamically such that the
black hole horizons remain centered on the excision boundaries.  As
described in Ref.~\cite{Scheel2006}, this is accomplished through a
control system that constantly monitors the location of the black
holes, and dynamically changes the functions $\lambda_\alpha(t)$
appropriately.  Such a control system is formulated most easily in
terms of control parameters $Q_\alpha\equiv\{Q_a, Q_\phi,Q_x,Q_y\}$
which have the properties (i) that $Q_\alpha=0$ when the black holes
are at their desired locations, and (ii) for small values of
$Q_\alpha$, changing the mapping-parameters $\lambda_\alpha$ changes
the control parameters $Q_\alpha$ according to
\begin{equation}\label{eq:ControlSystemOrthogonality}
\frac{\partial Q_{\alpha}}{\partial
\lambda_{\beta}}=-\delta^\alpha{}_\beta,\qquad\mbox{for }|Q_\alpha|\ll 1.
\end{equation}
The control parameters must be given in terms of the moving
coordinates of the centers of the apparent horizons, $c_{1,2}^i$, and
they must vanish when $c_{1,2}^i$ are at the desired locations,
namely, when they are at their values in the initial data
  $\left(c_{1,2}^i \right)_{t=0}$.  The derivatives in
Eq.~(\ref{eq:ControlSystemOrthogonality}) are to be taken at constant
inertial coordinates of the centers of the horizons.

To begin, we define 
\begin{align}
(\Delta_x(t),\Delta_y(t),\Delta_z(t))&\equiv\vec c_1(t) -\vec c_2(t),\\
D(t) &\equiv \big[\Delta_x^2(t)+\Delta_y^2(t)\big]^{1/2}.
\end{align}
Because of symmetries, $\Delta_z$ is always zero, and will not be
used.  The control parameters for the expansion factor $a(t)$ and
the rotation angle $\phi$ are given by
\begin{subequations}
\label{eq:ControlParameters}
\begin{align}
\label{eq:Qa}
Q_a&=a(t)\left(\frac{D(t)}{D_0}-1\right),\\
\label{eq:Qphi}
Q_\phi&=\frac{\Delta_y(t)}{D(t)}.
\end{align}
It is straightforward to verify that $Q_a$ and $Q_\phi$ satisfy
Eq.~(\ref{eq:ControlSystemOrthogonality}).

The control parameters for the translation are somewhat more involved.
We use the ansatz
\begin{equation}\label{eq:QT}
\left(\begin{aligned}
Q_x\\ 
Q_y
\end{aligned}\right)
=a(t)
\mathbf{R}_{\phi(t)}
\left[\left(
  \begin{aligned}
    x_B\\
    y_B
  \end{aligned}
\right)
+\mathbf{M}
\left(
  \begin{aligned}
    \Delta_x\\
   \Delta_y
  \end{aligned}
\right)
\right],
\end{equation}
\end{subequations}
where $\mathbf{M}$ is a constant $2 \times 2$ matrix, and we demand
that $\mathbf{M}$ commutes with $\mathbf{R}_{\phi(t)}$. Because
$\mathbf{M}$ and $\mathbf{R}_{\phi(t)}$ commute, Eq.~(\ref{eq:QT}) can
be rewritten in inertial coordinates as
\begin{equation}
\left(\begin{aligned}
Q_x\\ 
Q_y
\end{aligned}\right)
=
\left(
  \begin{aligned}
    \bar x_B\\
    \bar y_B
  \end{aligned}
\right)
+\mathbf{M}
\left(
  \begin{aligned}
    \bar\Delta_x\\
   \bar\Delta_y
  \end{aligned}
\right)
-
\left(
  \begin{aligned}
    T^{\bar x}\\
    T^{\bar y}
  \end{aligned}
\right),
\end{equation}
which makes it obvious that $Q_x$ and $Q_y$ satisfy
Eq.~(\ref{eq:ControlSystemOrthogonality}).  

To close this discussion, we must compute the matrix $\mathbf{M}$.
The requirements that $\mathbf{M}$ commute with $\mathbf{R}_\phi$ and
that $Q_x=Q_y=0$ for $c_{1,2}^i=\left(c_{1,2}^i \right)_{t=0}$
determine $\mathbf{M}$ uniquely:
\begin{equation}
\mathbf{M}=
\frac{1}{D_0}
\left(
\begin{aligned}
x_{A,0} && -y_{A,0} \\
y_{A,0} &&  x_{A,0}
\end{aligned}
\right).
\end{equation}

The mapping given in Eq.~(\ref{eq:SpatialCoordMap}) and the control
parameters, given in Eqs.~(\ref{eq:ControlParameters}), are then
combined with the feedback control system described in
Ref.~\cite{Scheel2006} in order to evolve the unequal-mass BBH through
the inspiral phase.

%%%%%%%%%%%%%%%%%%%%%%%%%%%%%%%%%%%%%%%%%%%%%%%%%%%%%%%%%%%%%%%%
\subsubsection{Constraint Damping}
\label{sec:ConstraintDamping}
%%%%%%%%%%%%%%%%%%%%%%%%%%%%%%%%%%%%%%%%%%%%%%%%%%%%%%%%%%%%%%%%

In order to suppress violations of the generalized harmonic gauge
constraint Eq.~(\ref{e:ghconstr})
(cf. Refs.~\cite{Pretorius2006,Gundlach2005}), and of the auxiliary
constraints that arise from the reduction of the generalized harmonic
evolution system to first order form
(cf. Ref.~\cite{Lindblom2006,Holst2004}), we introduce so-called {\em
  constraint damping terms} in the generalized harmonic evolution
equations (see \cite{Lindblom2006}).  These terms are proportional to
the {\em constraint damping parameters} $\gamma_0$ and $\gamma_2$.

Simulations with mass ratios $q=\{2,3\}$ were found to be stable with
the same constraint damping parameters as those used in
Ref.~\cite{Boyle2007}.  However, for the higher mass ratios
$q=\{4,6\}$, we encountered constraint violations that grew
exponentially on time scales of several $100M$.  We found that toward
the outer edges of the cylindrical subdomains, the constraint damping
parameters must be sufficiently large in order to suppress exponential
constraint growth.  In the overlap between the inner spherical shells
and the cylinders, an instability develops unless the constraint
damping is sufficiently {\em small}.  Furthermore, we were not able to
achieve stable evolutions with $\gamma_0=\gamma_2$.  After
considerable experimentation, we settled on a sum of Gaussians:
\begin{align}\label{eq:gamma0}
  M\gamma_0 =& 8 e^{-(r_1/1.3M)^2} + 16 e^{-(r_2/M)^2}+f_{\rm far-field}(r)\\
\label{eq:gamma2}
  M\gamma_2 =& 8 e^{-(r_1/1.3M)^2} + 40 e^{-(r_2/M)^2}+f_{\rm far-field}(r)
\end{align}
with far-field terms $f_{\rm far-field}= 0.2 e^{-(r/60M)^2} +
0.001$. Here $r_1$ and $r_2$ are the coordinate distances from the
centers of each hole, and $r$ is the distance from the origin. The
choices Eqs.~(\ref{eq:gamma0}) and~(\ref{eq:gamma2}) were found to
work well even for $q=\{2,3\}$, and all simulations presented here use
them.

We infer from these results that the domain decomposition with spheres
overlapping cylinders is not always stable, and that stability depends
sensitively on certain geometric details. Recent shorter simulations
that do not have overlapping subdomains do not show such
sensitivity. However, the domain decomposition of spheres and
cylinders is computationally more efficient, and therefore we employ
it during long inspiral simulations.

%%%%%%%%%%%%%%%%%%%%%%%%%%%%%%%%%%%%%%%%%%%%%%%%%%%%%%%%%%%%%%%%
\subsection{Eccentricity removal}
\label{sec:EccentricityRemoval}
%%%%%%%%%%%%%%%%%%%%%%%%%%%%%%%%%%%%%%%%%%%%%%%%%%%%%%%%%%%%%%%%

The procedure for eccentricity removal developed in
Refs.~\cite{Pfeiffer-Brown-etal:2007,Boyle2007} assumed an equal-mass
binary.  Generalization to unequal-mass binaries is straightforward.
As in Ref.~\cite{Boyle2007}, we fit the radial velocity (represented
by the time derivative of the proper separation $s(t)$ between the
horizons) by the functional form
\begin{equation}
\frac{ds}{dt}=v_{\rm insp}(t) + B \cos(\omega t+\phi).
\end{equation}
Here $v_{\rm insp}(t)$ is a monotonic
function varying on the (long) inspiral time scale; this function captures the
desired zero-eccentricity inspiral driven by radiation-reaction.  We
take here the functional form
\begin{equation}\label{eq:vinsp}
  v_{\rm insp}(t)=v_0 + v_1t + v_2 t^2,
\end{equation}
with three fitting parameters $v_0, v_1, v_2$.  However, in more
recent work~\cite{Buonanno:2010yk}, we describe fitting
functions that result in more robust behavior.  The oscillating piece
$B\cos(\omega t+\phi)$ captures superposed oscillation due to non-zero
orbital eccentricity -- the goal is to reduce the amplitude of this
piece.
%%%%%%%%%%%%%%%%%%%%%%%%%%%%%%%%%%%%%%%%%%%%%%%%%%%%%%%%%%%%%%%%
\begin{figure}
\includegraphics[width=0.9\columnwidth]{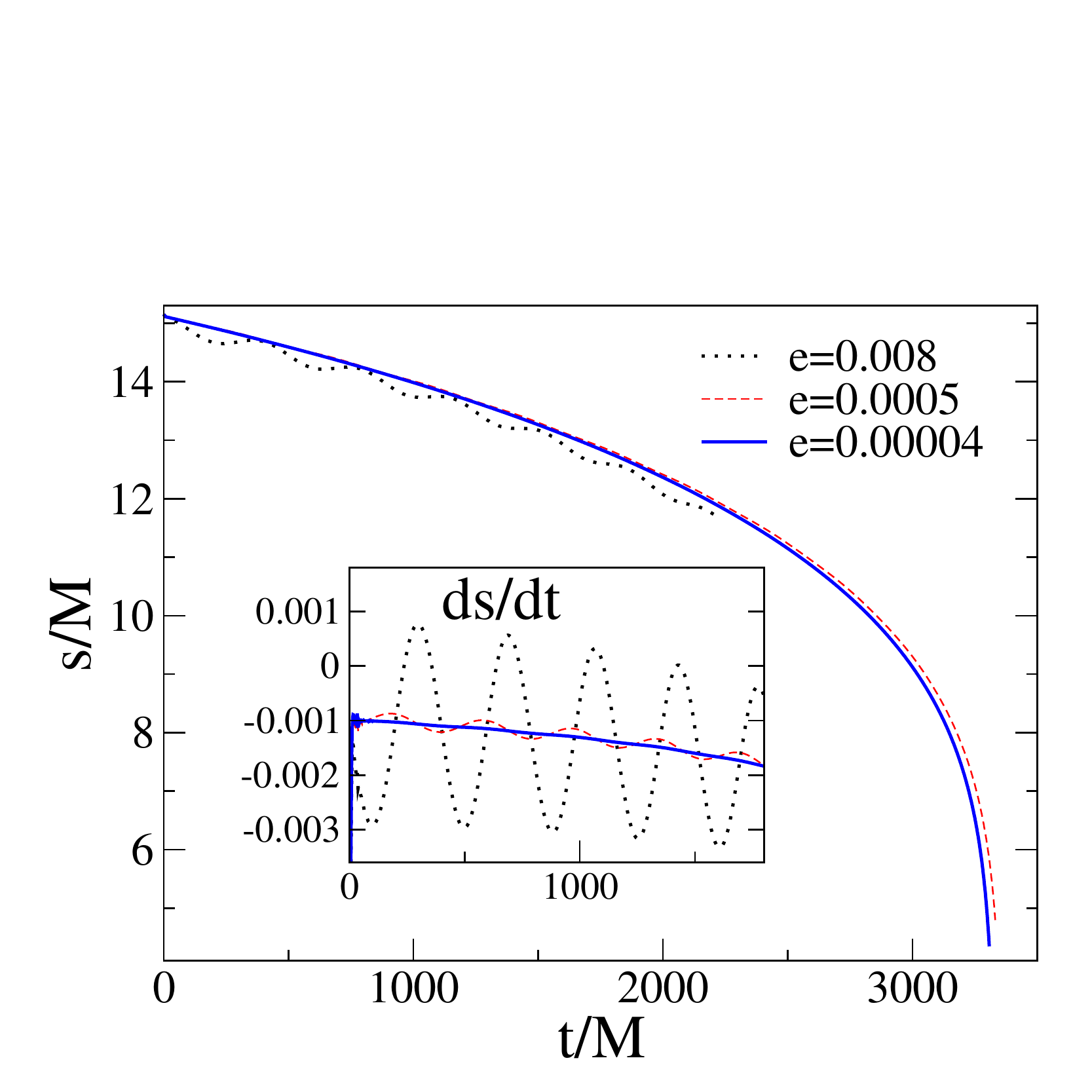}
\caption{\label{fig:q4_EccRemoval_PropSep} Eccentricity removal for
  mass ratio $q=4$.  The main panel shows proper separation as a
  function of time, the inset radial velocity $ds/dt$. Initial data
  parameters based on TaylorT3 post-Newtonian approximation result in the black
  dotted line with $e\approx 0.008$.  The red-dashed and the
  solid-blue lines represent two iterations of eccentricity removal,
  for a final eccentricity of $e\approx 4\times 10^{-5}$.}
\end{figure}
%%%%%%%%%%%%%%%%%%%%%%%%%%%%%%%%%%%%%%%%%%%%%%%%%%%%%%%%%%%%%%%%

For unequal masses, the black holes have different separations from the
origin, and therefore have different radial velocities.  To avoid
dealing with each black hole independently, we consider the initial
data specified in terms of a Hubble-like radial expansion factor
$\dot{a}_0$, which induces radial velocities proportional to the
distance to the origin, $v_r^i = \dot{a}_0\,x^i $ at a coordinate
location $x^i$. The updating formulas become
\begin{align}
\Omega_{0\, \rm new} &= \Omega_0  +\frac{B}{2s_0}\sin(\phi),\\
\dot{a}_{0\,\rm new} &= \dot{a}_0 -\frac{B}{s_0}\cos(\phi).
\end{align}
The orbital eccentricity is given by
\begin{equation}
e_{ds/dt}  = \frac{B}{s_0\omega},
\end{equation}
which is the same formula as for the equal-mass case.

Overall, eccentricity removal works as well here as for the equal-mass
cases considered previously.  Fig.~\ref{fig:q4_EccRemoval_PropSep}
shows that with each iteration, $e$ drops by about a factor of 10.
The most important factor for effective eccentricity removal is the
quality of the fit.  The fitting interval $[t_1,t_2]$ can start only
after transients due to junk radiation have decayed.  However, because
the fit is used to infer radial velocity and acceleration at time
$t=0$, the fitting interval needs to be sufficiently early in the run
to allow accurate extrapolation from the fitting interval back to
$t=0$.  Finally, the fitting interval needs to be long enough to allow
a reliable fit of the frequency $\omega$, i.e. it needs to be longer
than one period of the radial oscillations.  Inclusion of the term
quadratic in $t$ in Eq.~(\ref{eq:vinsp}) significantly improves the
quality of the fits and the effectiveness of the eccentricity removal.
For the runs described here, we choose $t_1$ on the order of $100M$
and $t_2$ on the order of $1000M$.

%%%%%%%%%%%%%%%%%%%%%%%%%%%%%%%%%%%%%%%%%%%%%%%%%%%%%%%%%%%%%%%%
\subsection{Evolution of merger \& ringdown}
\label{sec:MergerRingdown}
%%%%%%%%%%%%%%%%%%%%%%%%%%%%%%%%%%%%%%%%%%%%%%%%%%%%%%%%%%%%%%%%

The evolution algorithm for the inspiral described in
Section~\ref{sec:EvolutionInspiralPhase} fails when the black holes
approach each other too closely.  This failure is caused by several
factors. First, the gauge fields $H_a$ are chosen during inspiral to
be time-independent in the grid frame. This works well for the
inspiral because the solution (in the grid frame) is roughly
time-independent near the black holes. Near merger, however, this
gauge leads to the formation of coordinate singularities.  Second,
during inspiral, the excision boundaries of the grid remain spherical,
and do not change shape even though the individual apparent horizons
become distorted as the holes approach each other.  As the distortion
of the apparent horizons increases, the mismatch between the excision
boundaries and the apparent horizons eventually leads to a violation
of the excision condition, i.e., the condition that all characteristic
fields of the hyperbolic system are outgoing (i.e. into the hole) at
each excision boundary.  Third, the overlapping domain decomposition
used during the inspiral is prone to weak instabilities that cause no
trouble during the inspiral but drive rapidly growing modes after the
solution becomes highly dynamical.

%%%%%%%%%%%%%%%%%%%%%%%%%%%%%%%%%%%%%%%%%%%%%%%%%%%%%%%%%%%%%%%%
\begin{figure}
  \centerline{\setlength{\fboxsep}{0.5pt}%
\setlength{\fboxrule}{.8pt}%
\fbox{\includegraphics[width=\columnwidth]{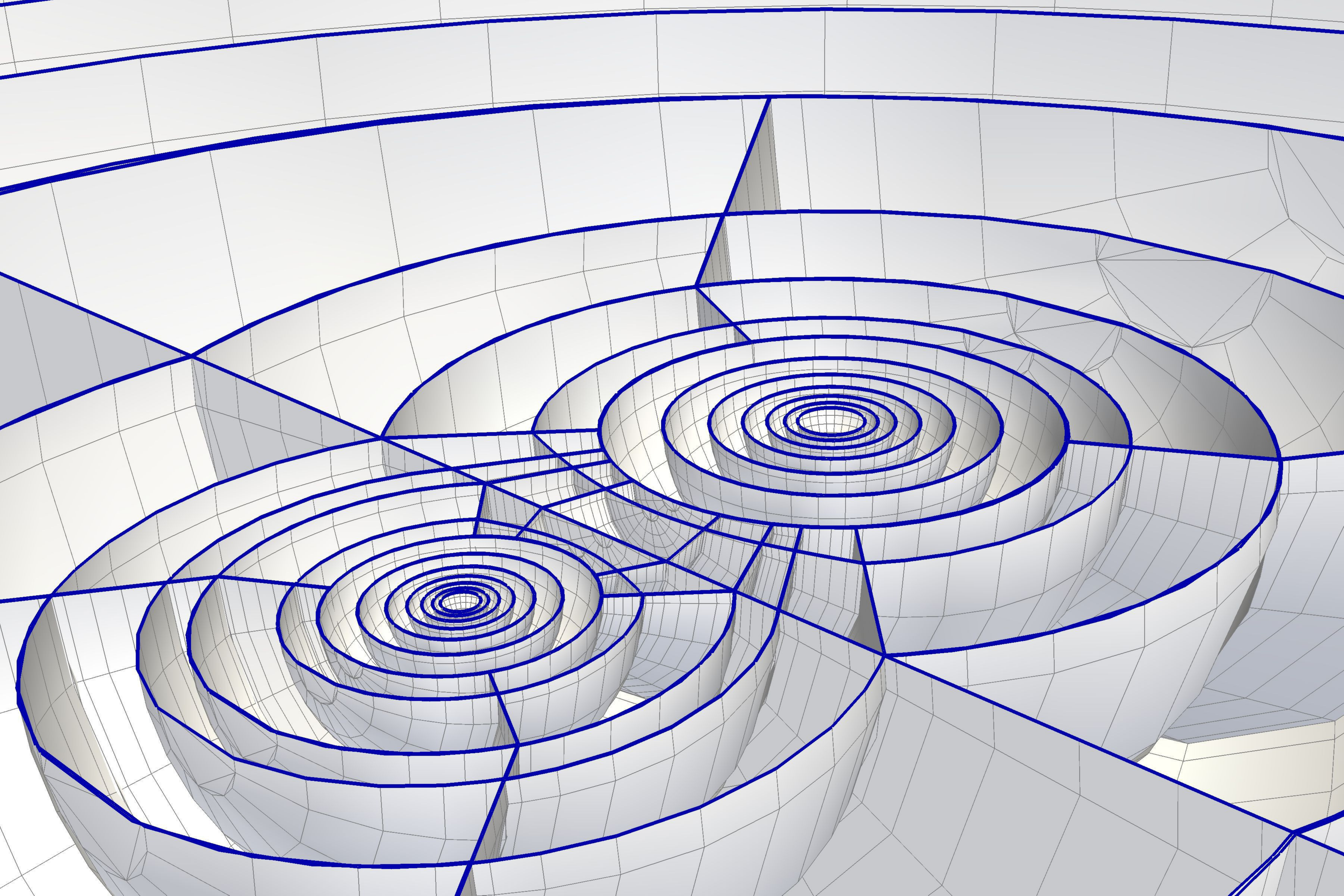}}}
  \caption{\label{fig:CutSphere3D} Domain decomposition used
for the plunge and merger for mass ratio $q=2$.
The thick blue lines represent subdomain boundaries in the $z\!=\!0\,$ plane.
The region $z\!>\!0$ is not shown. Also not shown is the additional 
deformation of the grid near the black holes that matches the shape of the
excision spheres to the apparent horizons.}
\end{figure}
%%%%%%%%%%%%%%%%%%%%%%%%%%%%%%%%%%%%%%%%%%%%%%%%%%%%%%%%%%%%%%%%

To address these problems, we stop the simulation about 1.5 orbits
before merger, and restart with a modified algorithm.  We change
smoothly to a damped harmonic
gauge~\cite{Lindblom2009c,Choptuik:2009ww,Szilagyi:2009qz} that slows
down the formation of coordinate singularities.  We also dynamically
modify the coordinate mapping between the grid frame and the inertial
frame so that the excision boundaries conform to the shapes of the
apparent horizons~\cite{Scheel2009,Szilagyi:2009qz}.  Furthermore, by
monitoring the characteristic speeds of the system, we dynamically
vary the velocity (with respect to the horizon) of each excision
boundary so as to ensure that the characteristic fields are outgoing
at these boundaries for all times; this characteristic speed control
is also crucial for evolving BBHs with large
spins~\cite{Lovelace:2010ne}.  Finally, we run the simulation on a set of
non-overlapping subdomains consisting of topological cubes,
cylindrical shells, and spherical shells.  This domain
decomposition is shown in Fig.~\ref{fig:CutSphere3D}.
 Each subdomain is distorted
by a coordinate mapping so that the subdomains do not overlap and so
that the union of these subdomains covers the entire 3-dimensional
region (minus two excised holes) inside a spherical outer boundary
$R_{\mathrm Bdry}$ of order a few hundred $M$ from the source (see
Section~\ref{sec:outerboundary} where we compare runs with different
values of $R_{\mathrm Bdry}$).   More details about the merger domain decomposition are given in the Appendix.  It avoids certain instabilities
that appear for domain decompositions with overlapping grid close to
merger~\cite{Szilagyi:2009qz}.  In addition, we choose a slightly
higher resolution for the non-overlapping grid than for the
overlapping grid used during inspiral, because the merger has features
with a shorter length scale than in the inspiral.  After the binary
has reached about $t\sim 2M$ before merger, we increase the resolution
one last time, particularly in the region between the two
holes.\footnote{The processes of regridding, changing resolution, and
  changing the coordinate mapping have since been automated; this will
  be described in a future work.}

After a common apparent horizon forms, we regrid onto a new set of
subdomains consisting of nested distorted spherical shells. The
innermost boundary is just inside the common apparent horizon, and
conforms to its shape.  The outermost boundary is the same $R_{\mathrm
  Bdry}$ used in the merger. The matching of the ringdown to the
inspiral is discussed in~\cite{Szilagyi:2009qz}.

\subsection{Relation to other {\tt SpEC} simulations}

\label{step:CommonCode}

Several other {\tt SpEC} simulations of binary black holes have been
presented in the
literature~\cite{Lovelace:2011nu,Lovelace:2010ne,Chu2009,%
  Scheel2009,Boyle2007}.  In this section we briefly describe some
computational details common to all {\tt SpEC} simulations, and we
describe how some of the new computational infrastructure presented
here relates to these other simulations.

Our apparent horizon finder expands the radius of the apparent horizon
as a series in spherical harmonics up to some order $L$.  We utilize
the fast flow methods developed by Gundlach~\cite{Gundlach1998} to
determine the expansion coefficients.  The quasi-local spin $S$ of
each black hole is computed with the spin diagnostics described
in~\cite{Lovelace2008}.  We compute the spin from an angular momentum
surface integral~\cite{BrownYork1993,Ashtekar-Krishnan:2004} using
approximate Killing vectors of the apparent horizons, as described
in~\cite{OwenThesis,Lovelace2008} (see
also~\cite{Dreyer2003,Cook2007}).  We define the dimensionless spin by
\begin{equation}\label{eq:SMMDef}
\chi=\frac{S}{M^2}.
\end{equation}

We extract gravitational waves from our simulations by two independent
methods.  We compute the Newman-Penrose scalar $\Psi_4$ using the same
procedure as described in~\cite{Pfeiffer-Brown-etal:2007,Boyle2007}.
This involves constructing the correct contraction of the Weyl
curvature tensor at several finite-radius coordinate-spheres far from
the source and projecting into spin-weighted spherical harmonics.  We
also extract the Regge-Wheeler-Zerilli
(RWZ)~\cite{ReggeWheeler1957,Zerilli1970b} gravitational wave strain
$h_{\ell m}$ as formulated in Ref.~\cite{Sarbach2001}. The
implementation of this formulation in the {\tt SpEC} code is described
in~\cite{Rinne2008b} (see also~\cite{PanEtAl:2011} and the appendix
of~\cite{Buonanno:2009qa} for further details). Both the $\Psi_4$ and
the RWZ waveforms, which are extracted at a series of finite-radius
coordinate spheres, are extrapolated to infinite distance from the
source~\cite{Boyle-Mroue:2008}.  The $\Psi_4$ waveforms generally
agree well with the (second time derivative of the) RWZ $h_{\ell m}$
waveforms, although for some purposes RWZ is a better choice than
$\Psi_4$ or vice versa. For example, computing strain from $\Psi_4$
requires two time integrations and careful choice of integration
constants, so it is simpler and less error-prone to instead use RWZ to
compute strain.  Similarly, computing the recoil velocity requires
either a time derivative of $h_{\ell m}$ or a time integral of
$\Psi_4$; the time derivative amplifies noise in the waveform, and
this affects the recoil velocity enough that it is better to use a
time integral of $\Psi_4$ for that purpose.

In parallel to the present work, superposed Kerr-Schild initial
data\cite{Matzner1999,Pfeiffer2003a,Lovelace2008} have been developed
and applied to {\tt SpEC} simulations of black holes with high
spins~\cite{Lovelace:2011nu,Lovelace:2010ne}.  The algorithmic
improvements discussed in the present work are generally compatible
with superposed Kerr-Schild simulations.  Specifically, the
root-finding procedure discussed in Sec.~\ref{sec:ID} can be applied
to superposed Kerr-Schild initial data.  This requires a change of
free parameters from excision sphere radii to masses of the conformal
black holes in the superposed Kerr-Schild initial data.  Early tests
indicate that the root-finding procedure works satisfactorily.
However, more exhaustive tests, especially for high spin systems, will
be necessary.

The control system discussed in Sec.~\ref{sec:CoordMap} is applicable
to any non-precessing simulation, independent of the type of initial
data.  The choice of gauge source functions $H_a$ (equal to the values
in the initial data, with appropriate coordinate transformations
applied~\cite{Boyle2007,Scheel2009,Chu2009}) does not work for
simulations with moderate or large spins; such simulations use active
gauge conditions already during the inspiral, see
e.g.~\cite{Lovelace:2011nu}.  Furthermore, moderate to high spin
simulations require use of a non-overlapping domain decomposition
during the inspiral to avoid certain grid instabilities.  We have no
reason to believe that the more complex and computationally more
expensive technology for high-spin systems might fail for the present
non-spinning simulations.  We have not tested this, because the
methods presented here are more efficient for the systems being
studied here.  Techniques for handling the merger, as described in
Sec.~\ref{sec:MergerRingdown} and the Appendix, are common between the
high-spin simulations and the simulations presented here.

%#########################
\section{Results}
\label{sec:Results}
%#########################

\subsection{Overview}
\begin{figure}
\vspace*{-1em}\includegraphics[width=1.0\columnwidth]{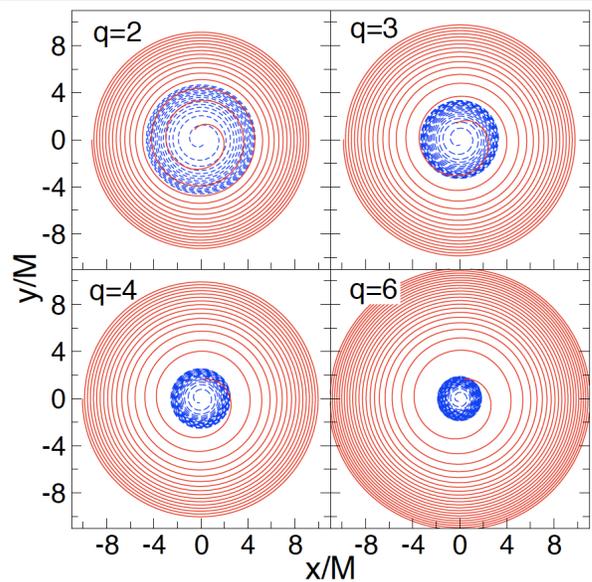}
\caption{Orbital trajectories for mass ratio $q=2,3,4,6$.
  For all mass ratios,
    the trajectory of the larger hole is represented by a dashed blue
    line, and that of the smaller hole by a solid red line.}
\label{Fig:Trajectories}
% this is with new data
\end{figure}

In this section, we present the results of our simulations of
non-spinning binary black holes with mass ratios $q=2,3,4,6$. These
simulations contain long inspirals (15 to 22 orbits), merger, and
ringdown. To achieve our desired number of inspiral orbits, we compute
the initial coordinate separation $D_0$ using Taylor T3 post-Newtonian
predictions~\cite{Blanchet2006}, and then proceed to the eccentricity
removal procedure as explained in
Sec.~\ref{sec:EccentricityRemoval}. Our final parameters for the
initial data set are summarized in Table~\ref{Table:RunParameters},
and Fig.~\ref{Fig:Trajectories} shows the trajectories of all our runs
through inspiral, the formation of a common apparent horizon, and
merger.

\begin{table*}
\begin{tabular}{|c||c|c|c|c|c||c|c||c|c||l|l|r|} 
  \hline 
  $\;q\;$   &\multicolumn{5}{|c||}{Initial Data}
&\multicolumn{2}{|c||}{ $R_{\rm Bdry}$}
&\multicolumn{2}{|c||}{Inspiral}
&\multicolumn{3}{|c|}{Remnant properties} 
%& $M/M_{\rm initial}$ & $\;\;\;S/M^2$
\\  
  &$10^3M\Omega_0$&$10^6\dot{a}_0M$&$D_0/M$ & $E_{\rm ADM}/M$ & $J_{\rm ADM}/M^2$ 
& $t=0$ 
& $t\to\mbox{late}$ 
&$10^5\varepsilon_{ds/dt}$&$N_{\rm GW}$
&$M_{{\rm c}, f}/M$&$S_f/(M_{{\rm c}, f})^2$&$v_{\rm kick}{\rm(km/s)}$ \\
  \hline
%
% 0093g  Boyle et al 2007, Scheel et al 2008
  1   & 16.7081  &  -28.40 & 14.4363 & 0.992333 & 1.0857 & 460M & 290M 
& 5 & 33 & 0.95162(2) & 0.68646(4) & 0 \\ \hline 
%
% /mnt/raid-project/nr/pfeiffer/ProductionRuns/q02_s0_s0/Final/Ev2
  2   & 17.6711  &  -62.53 & 13.8738 & 0.993025 & 0.9555 & 444M & 442M 
& 3 & 31 & 0.96124(2) & 0.62344(4) & 148(2)\\ \hline 
%
% /mnt/raid-project/nr/pfeiffer/ProductionRuns/q03_s0_s0/Final/Ev2
  3   & 18.9994  &  -63.63 & 13.1767 & 0.993868 & 0.7922 & 422M & 420M 
& 2 & 31 & 0.97128(1) & 0.54058(2) & 174(6)\\ \hline 
%
% /mnt/raid-project/nr/pfeiffer/ProductionRuns/q04_s0_s0/Final/Ev2
  4   & 20.3077  &  -66.08 & 12.5652 & 0.994568 & 0.6655 &402M & 400M 
& 4 & 31 &   0.97792(2) & 0.47160(10) & 157(2) \\ \hline
%
% /mnt/raid-project/nr/pfeiffer/ProductionRuns/q06_s0_s0/d13/Final/Ev9
  6    & 19.35244 & -42.43 & 13.0000 & 0.995968 & 0.5157 & 572M & 569M 
& 4 & 43 &  0.98547(5) & 0.37245(10) & 118(6)\\ \hline 
%
% S025a on Sugar, initialdata at /home/harald/research/BBH/SpECID/08Jun15/Ev1
%  6   & 29.5914  &  -150.1 & 9.58293 & 0.994881 & 0.4684 & 307M & 265M 
%&8 &  16(?)  & 0.98547(1) &0.37245(1)  \\ \hline 

\end{tabular} 
\caption{\label{Table:RunParameters} Runs considered in this paper,
  with $q=1$ from Ref.~\cite{Scheel2009} included for completeness.
  Initial data parameters are orbital frequency $\Omega_0$, the
  expansion factor $\dot{a}_0$, and the coordinate distance between the
  black hole centers $D_0$.  Furthermore, the initial and final radii
  of the outer boundary are given ($R_{\rm Bdry}$ is decreasing during
  the evolution, cf.~\cite{Boyle2007}), as well as the initial orbital eccentricity
  $\varepsilon_{ds/dt}$ and the number of gravitational wave (GW)
  cycles before the peak of $|h_{22}|$, $N_{\rm GW}$. The last three
  columns denote the Christodoulou mass, dimensionless spin, and kick
  velocity of the merged black hole at the end of ringdown. 
% Mfinal, Sfinal from Marks April 27, 2011 email 
}
\end{table*}

%%%%%%%%%%%%%%%%%%%%%%%%%%%%%%%%%%%%%%%%%%%%%%%%%%%%%%%%%%%%%%%%
\subsection{Mass calibration}
\label{SubSec:MassRelxn}
%%%%%%%%%%%%%%%%%%%%%%%%%%%%%%%%%%%%%%%%%%%%%%%%%%%%%%%%%%%%%%%%

\begin{figure}
\includegraphics[width=0.99\columnwidth]{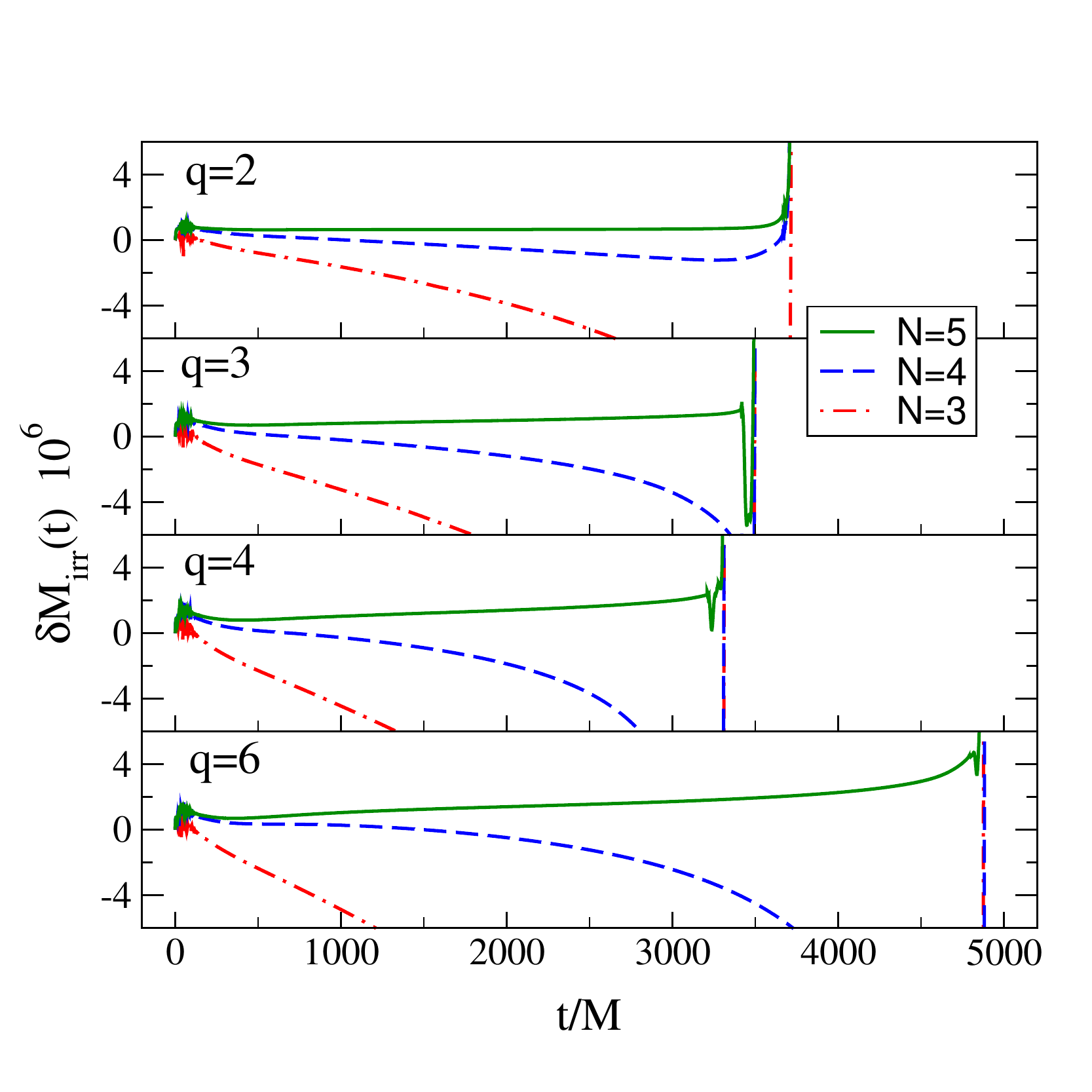}
\caption{Convergence of the irreducible mass.  Plotted is $\delta
  M_{\rm irr}\equiv \left[M_{\rm irr}(t)-M_{\rm irr}(0)\right] /
  M_{\rm irr}(0)$ for three different numerical resolutions ($N=3,4,5$).}
\label{Fig:MassRelxn}
% this is with new data
\end{figure}

A mass scale $M$ by which all data are rescaled is defined as
follows. Consider the sum of the two irreducible masses, defined from
the areas $A_{{\rm AH}\,1}$ and $A_{{\rm AH}\,2}$ of the apparent horizons,
\begin{equation} \Mirr(t) \equiv
  \sqrt{\frac{A_{{\rm AH}\,1}(t)}{16\pi}}
  \;+\;\sqrt{\frac{A_{{\rm AH}\,2}(t)}{16\pi}}. 
\end{equation}
Root-finding during construction of the initial data ensures
$\Mirr(0)=1$.  Figure~\ref{Fig:MassRelxn} presents convergence data
for the irreducible mass during the simulations.  Plotted is the
relative change of $\Mirr(t)$.  Convergence is clearly apparent, and
the irreducible mass is constant to within a few parts in $10^6$ at
the highest resolution, except immediately before merger.  During the
first $\sim 100M$, the black hole mass increases by about $1\times
10^{-6}$.  Since this is below the numerical error during inspiral
shown in Fig.~\ref{Fig:MassRelxn}, we define our mass scale by
\begin{equation}
  M\equiv M_{\rm irr}(0)
\end{equation}
for all mass ratios.

%%%%%%%%%%%%%%%%%%%%%%%%%%%%%%%%%%%%%%%%%%%%%%%%%%%%%%%%%%%%%%%%
\subsection{Accuracy}
%%%%%%%%%%%%%%%%%%%%%%%%%%%%%%%%%%%%%%%%%%%%%%%%%%%%%%%%%%%%%%%%

%%%%%%%%%%%%%%%%%%%%%%%%%%%%%%%%%%%%%%%%%%%%%%%%%%%%%%%%%%%%%%%%
\subsubsection{Phase convergence}
%%%%%%%%%%%%%%%%%%%%%%%%%%%%%%%%%%%%%%%%%%%%%%%%%%%%%%%%%%%%%%%%

One of the goals of the present work is to calculate long, accurate
waveforms for the dominant and top subdominant gravitational wave
modes -- $(2,2)$, $(3,3)$, and $(2,1)$ -- from unequal-mass binary
black hole simulations. The top subdominant modes are those with the
largest peak strain amplitude.  To determine the accuracy of these
waveforms, we perform convergence studies of RWZ-$h_{\ell m}$ at
a particular extraction radius.

All simulations are run at three different resolutions, labeled
$N=3,4,5$.  For all three resolutions, the RWZ gravitational
waveforms at a finite extraction radius ($R_{\rm ext}=338M$ for
$q=2,3,4$ and $R_{\rm ext}=460M$ for $q=6$) are computed.  We
decompose the complex spherical harmonic modes into real-valued
amplitude and phase:
\begin{equation} 
  h_{lm}(t) = A_{lm}(t) \exp(i\phi_{lm}(t)).
\end{equation}
We next compute differences $\Delta\phi_{lm}(t)$ between different
resolutions {\em without} any time shifts,
\begin{equation}
  \Delta\phi^{N\,N'}_{lm}(t) = \phi^N_{lm}(t) - \phi^{N'}_{lm}(t),
\end{equation}
where the superscripts $N$ and $N'$ refer to the numerical resolutions
being considered.  Finally, for ease of presentation, we time-shift
the phase differences to align convergence tests of different mass
ratios at their respective times of peak amplitude of the $h_{22}$
mode, $t_{{\rm peak}\, 22}$.

\begin{figure}
\includegraphics[scale=0.50]{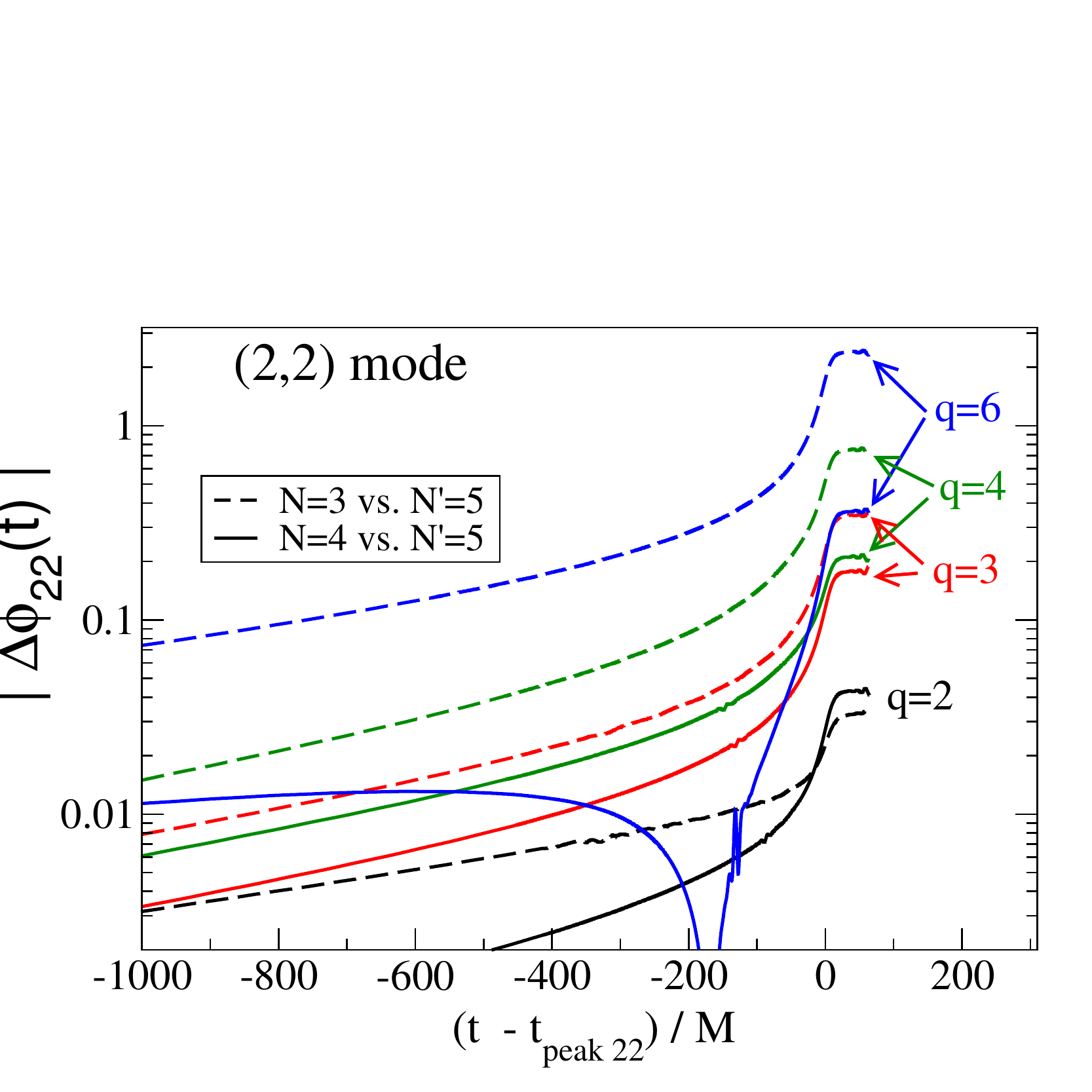}
\caption{Phase convergence of RWZ-h (2,2) modes for
  inspiral-merger-ringdown waveforms. Shown are the phase differences
  between a given resolution and the highest resolution.  }
\label{Fig:GWPhaseConv22}
\end{figure}

Phase differences for the dominant $(2,2)$ mode are plotted in
Fig.~\ref{Fig:GWPhaseConv22}.  Note that this figure shows only the
part of the simulation around merger time.  During the earlier
inspiral, the phase errors are lower. It is apparent from this plot
that the phase accuracy deteriorates with increased mass ratio, albeit
quite slowly.  This is expected, as simulations become numerically
more difficult with increased mass ratio, owing to the smaller
gravitational wave (GW) flux, and the smaller length scale of the
small black hole.  Nevertheless, the phase accuracies of all the new
simulations presented in this paper are comparable to that of the
equal-mass, zero spin simulation presented in Scheel et
al~\cite{Scheel2009}, with the simulations at low mass ratios
($q\!=\!2$) being somewhat more accurate, and those at higher mass
ratios ($q=3,4,6$) somewhat less accurate.  

Note that during merger
and ringdown, the three resolutions of the $q=2$ simulation do not
follow the usual pattern indicating convergence.  There are
a few possible reasons for this.  One is that for $q\!=\!2$, the 
truncation error as a function of resolution may change sign near 
one of the resolutions $N=3$, $4$, or $5$, thus producing an artificially
small truncation error and skewing the test shown in 
Fig.~\ref{Fig:GWPhaseConv22}.  Another possibility is
that the unusual pattern is caused by small differences in gauge or domain
decomposition between different resolutions: as explained in 
Section~\ref{sec:MergerRingdown}, 
we change the gauge and domain decomposition about 1.5 orbits before merger, 
but these changes occur at slightly different times for
different resolutions, and this time offset will introduce a small 
non-convergent
error.  Note also that the $q\!=\!2$ case appears to have 
a factor of three smaller truncation error than any previous long {\tt SpEC}
simulation, so this case may reveal small error sources 
that may not have been evident in previous simulations.
Figure~\ref{Fig:GWPhaseConv22} shows a feature in the $q=6$
  simulation around $t\sim -180M$.  This arises because the
  phase difference between $N=4$ and $N=5$ simulations changes sign.

\begin{figure}
\includegraphics[scale=0.50]{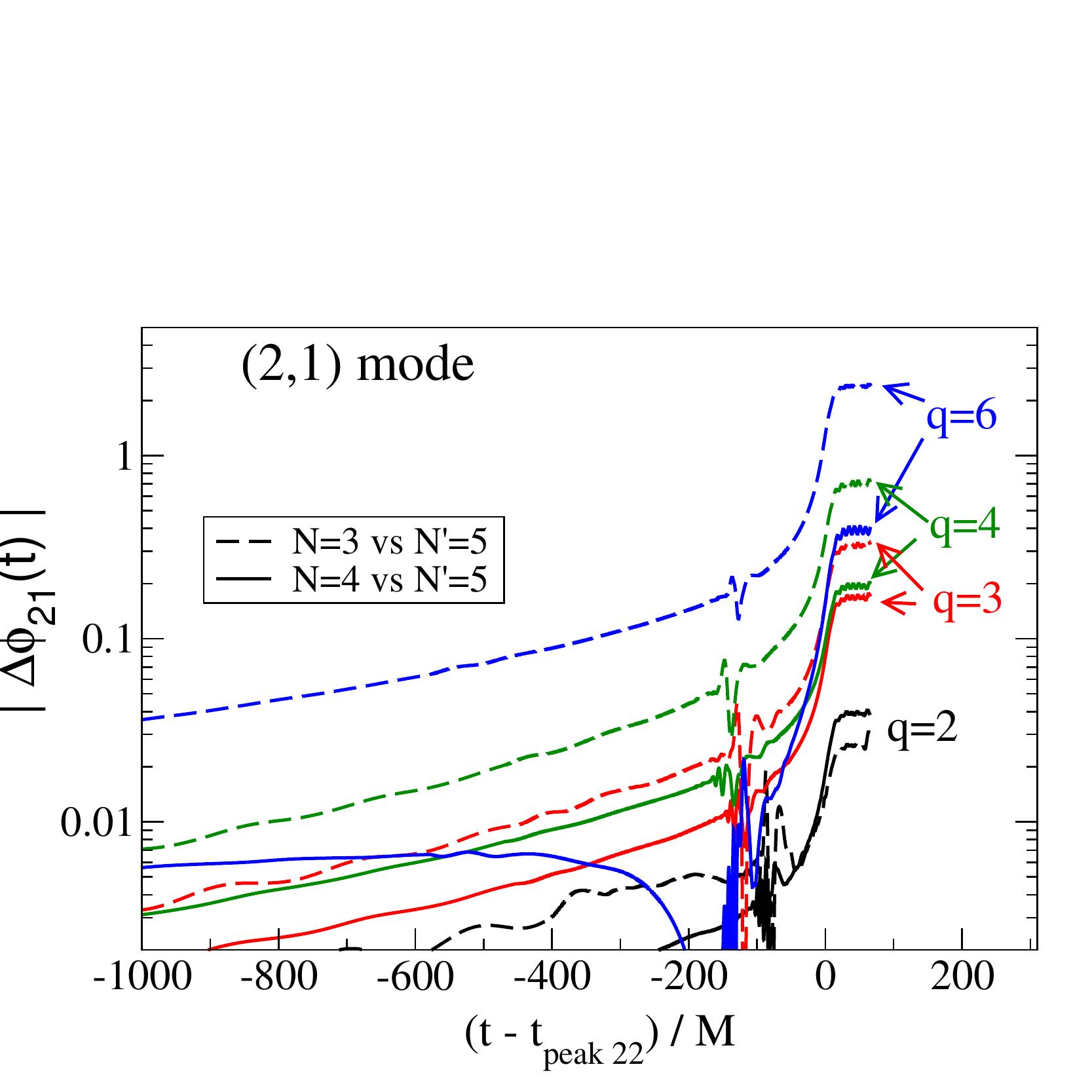}
%\\
\includegraphics[scale=0.50]{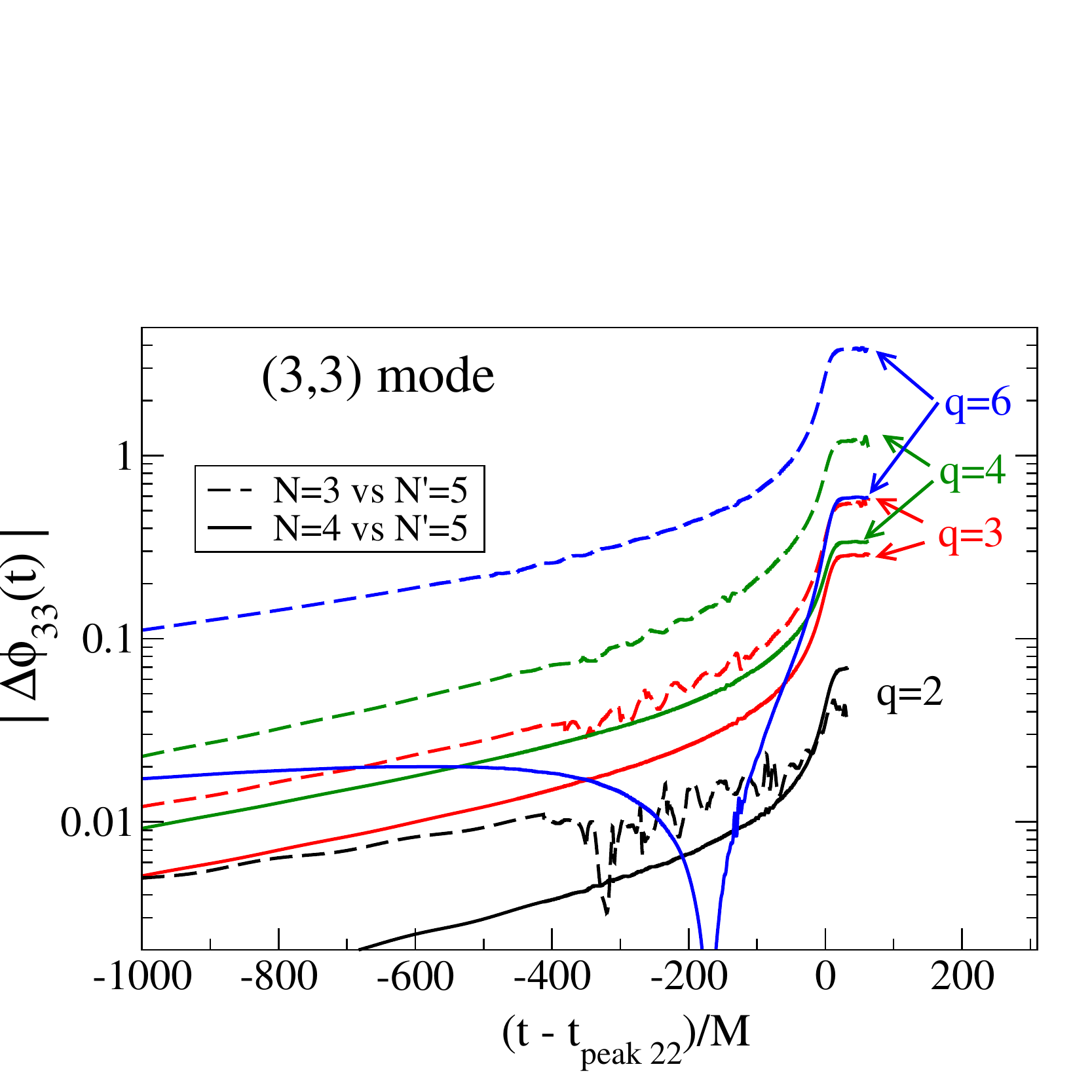}
\caption{As Fig.~\ref{Fig:GWPhaseConv22}, but for the subdominant
  modes (3,3) and (2,1).}
\label{Fig:GWPhaseConv33-21}
% this is with new data
\end{figure}

Convergence tests for the two leading subdominant modes $(2,1)$ and
$(3,3)$ are presented in Fig.~\ref{Fig:GWPhaseConv33-21}.  During the
inspiral, the phase errors of the (2,1) mode are approximately half as
large as those for the (2,2) mode, whereas the errors in the (3,3)
mode are approximately a factor $1.5$ larger.  This scaling is
reasonable, as all three GW modes are determined primarily by the
orbital phase evolution.  The gravitational wave mode $(l,m)$ proceeds
through $m$ cycles for each orbit; hence, the GW phase errors of
different modes should be proportional to $m$.  During merger and
ringdown, the observed phase errors behave differently:
$\Delta\phi_{33}$ is larger than $\Delta\phi_{22}$ for all mass
ratios, whereas $\Delta\phi_{21}$ is similar in amplitude to
$\Delta\phi_{22}$.  Fig.~\ref{Fig:GWPhaseConv33-21} shows noise in the
(2,1) convergence test, starting about $150M$ before peak amplitude.
Presumably, the noise in the phase is more prominent in the (2,1) mode
because of the small amplitude of this mode.

%%%%%%%%%%%%%%%%%%%%%%%%%%%%%%%%%%%%%%%%%%%%%%%%%%%%%%%%%%%%%%%%
\subsubsection{Effect of location of outer boundary}
\label{sec:outerboundary}
%%%%%%%%%%%%%%%%%%%%%%%%%%%%%%%%%%%%%%%%%%%%%%%%%%%%%%%%%%%%%%%%

The simulations presented here are of such long duration that the
black holes are in causal contact with the outer boundary for a large
portion of the evolution.  The question therefore arises: are the
results affected by our choice of outer boundary conditions? Ideally,
the gravitational waveforms computed on a truncated computational
domain with an artificial outer boundary should not have errors
introduced by the boundary conditions themselves -- either from
spurious reflections of gravitational radiation or from constraint
violations at the outer boundary. The extent to which this is achieved
indicates the degree to which the outer boundaries are ``absorbing''
(see e.g. Refs.~\cite{Buchman2006,Buchman2007, Rinne2007,
  Rinne2008b}).  The outer boundary conditions used in our simulations
are (i) constraint-preserving and (ii) freeze the Weyl scalar $\Psi_0$
to its initial value.  These ``semi-absorbing'' boundary conditions
are the simplest in a hierarchy of increasingly absorbing boundary
conditions, described in detail in Sec. 4.2 of~\cite{Buchman2006}.

\begin{table}[b]
\begin{tabular}{|c||c|c|c||c|}
\hline $\;q\;$ & $\;R_{\rm close}$ & $\;R_{\rm normal}$ & $\;\;R_{\rm far}\;\;$ &   
%$\frac{|A\,\gamma_2|_{21}}{|A\,\gamma_2|_{22}}$ 
$Q_{m=1,m=2}$
\\ \hline
2 & $20D_0$ & $32D_0$ & $50D_0$ & 0.54\\
3 & $20D_0$ & $32D_0$ & $50D_0$ & 0.80 \\
4 & $20D_0$ & $32D_0$ & $50D_0$ & 1.02 \\
6 & $26D_0$ & $44D_0$ & $74D_0$ & 1.14 \\ \hline
\end{tabular}
\caption{\label{Table:ReflCoeff} Radii of the outer boundary for the
  runs with different outer boundary locations (in units of the initial
  separation $D_0$).  Also given is the ratio $Q$ of spurious reflections
  from the (2,1) mode relative to those from the (2,2) mode, cf. Eq.~(\ref{eq:ReflectionRatio}).}
\end{table}

\begin{figure}
\centerline{  \includegraphics[width=0.9\columnwidth]{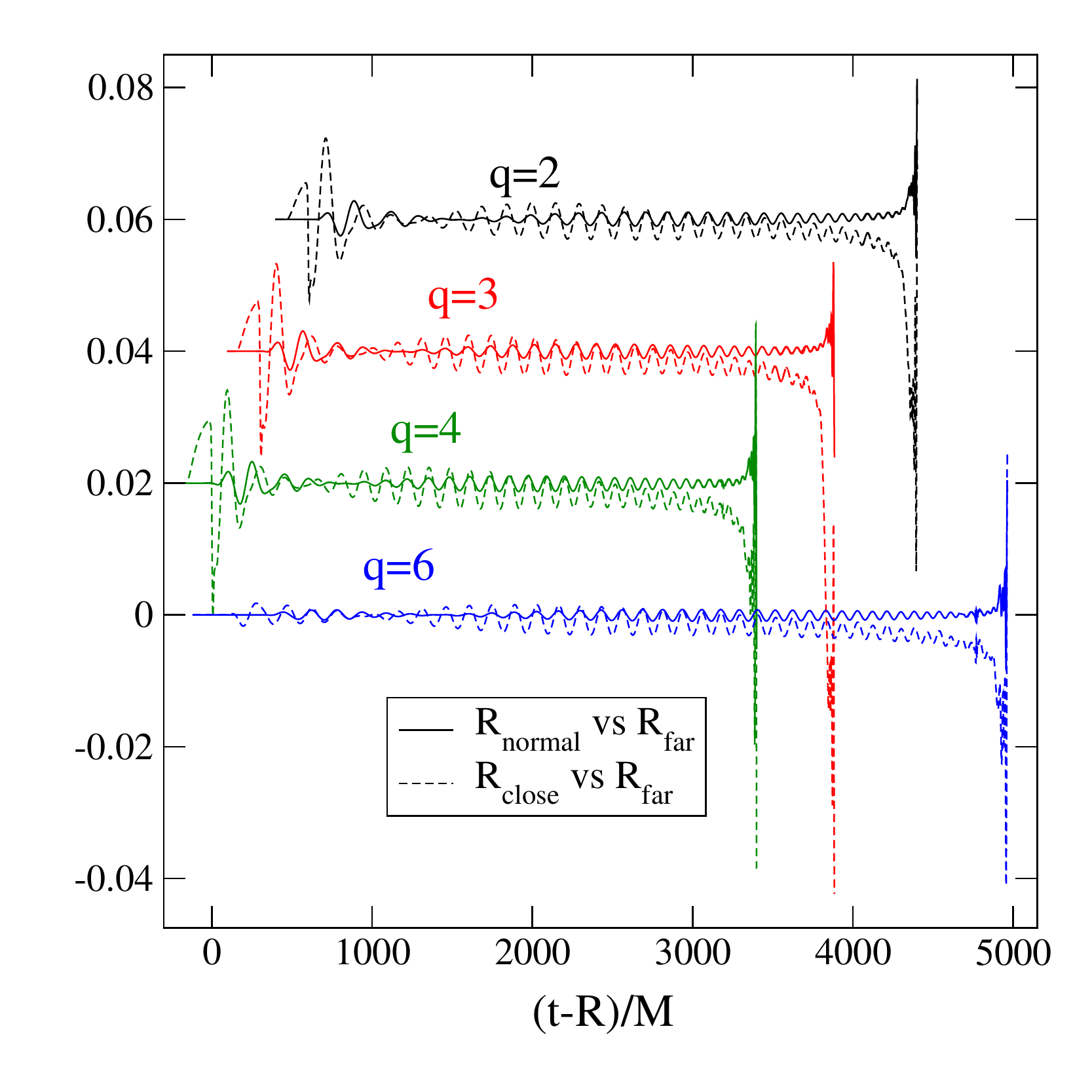}}
  \caption{Effect of the outer boundary location.  Shown are
    phase differences of $h_{22}$ between simulations with outer
    boundary radii given in Table~\ref{Table:ReflCoeff}.  The solid
    lines give the differences between ``normal'' and ``far''
    boundaries; the dashed lines give the differences between
    ``close'' and ``far'' boundaries.  For clarity, $q=2,3,4$ are
    offset vertically by multiples of $0.02$rad and $q=2,3$ are offset
    horizontally by multiples of $300M$.  $R$ denotes the extraction
    radius of each simulation. }
  \label{Fig:OuterBdryTests}
\end{figure}

To evaluate the impact of the artificial outer boundary on our
simulations, we repeat the $N=4$ simulations for each mass ratio with
two additional outer boundary radii, $R_{\rm close}$ and $R_{\rm
  far}$, where the distance to the outer boundary is changed {\em
  only} by adding or removing outer spherical shells in our domain
decomposition.  The different outer boundary radii are listed in
Table~\ref{Table:ReflCoeff}.  The $h_{22}$ waveforms are extracted from these
simulations, and phase differences between runs with different outer
boundary radii are computed and plotted in
Fig.~\ref{Fig:OuterBdryTests}.  The plotted phase differences are
oscillatory during inspiral, indicating that the runs being compared
have slightly different orbital eccentricities.  Around merger, a
systematic phase difference appears of a few times $0.01$rad for the
near boundary and $\lesssim 0.005$rad for the normal boundary
location.  During ringdown, the gravitational wave amplitude decays
exponentially and the calculation of the phase becomes increasingly
noisy.  We truncate the plotted data when the amplitudes of the waves
have decayed to 1\% of their peak values. It is evident from
Fig.~\ref{Fig:OuterBdryTests} that the transparency of the outer
boundary diminishes as the distance to the boundary decreases.  For
our ``normal'' boundary radius, the phase error due to the boundary is
$\lesssim 0.005$rad (when compared to the far location), which is
negligible relative to the truncation error presented in
Fig.~\ref{Fig:GWPhaseConv22}.  On the other hand, moving the boundary
from the ``normal'' to the ``close'' location increases phase errors 5
to 10 times.

We can relate the phase errors reported in
Fig.~\ref{Fig:OuterBdryTests} to the expected reflection coefficients
of our semi-absorbing boundary conditions as analyzed in
Ref.~\cite{Buchman2006}.  The quadrupolar wave $(\ell=2)$ reflection
coefficient $\sigma_2$ for freezing-$\Psi_0$ plus constraint preserving boundary
conditions is given by Eq.~(89) of Ref.~\cite{Buchman2006}.  In the
limit of large boundary radius $kR_{\rm Bdry}\gg 1$ (where $k$ is the
wavenumber of the outgoing wave), the reflection coefficient reduces
to
\begin{equation} \sigma_{2} =
  \frac{3}{2}\left(k R_{\rm Bdry}\right)^{-4}.\label{eq:sigma2}
\end{equation}
``Near'' boundaries are a factor $\sim 1.6$ closer than ``normal''
boundaries; therefore, the reflection coefficient will be larger by a
factor $1.6^4\approx 6.5$, consistent with the observed increase of
phase errors by a factor 5--10 in Fig.~\ref{Fig:OuterBdryTests}.
Moreover, according to an argument given in Ref.~\cite{Boyle2006}, the
phase error due to reflection of the $(2,2)$ mode of the outgoing
radiation should be roughly equal to $\sigma_{2}$ times the total
accumulated phase\footnote{Depending on assumptions, $\sigma_{2}$ may be
  raised to a power close to unity, cf. Eq.~(17) of
  Ref.~\cite{Boyle2006}.}.  For the $q=2,3,4$ simulations with normal
boundary locations, we have $kR_{\rm Bdry}\sim 18$ and $\sigma_{2}\sim
1.3\times 10^{-5}$.  The $\sim 30$ GW-cycles of inspiral correspond to
$\phi_{22}\sim 200$rad, 
so that $\sigma_{2}\phi_{22}\sim 0.003$rad, in
broad agreement with Fig.~\ref{Fig:OuterBdryTests}.

\begin{figure*}
\includegraphics[scale=0.7]{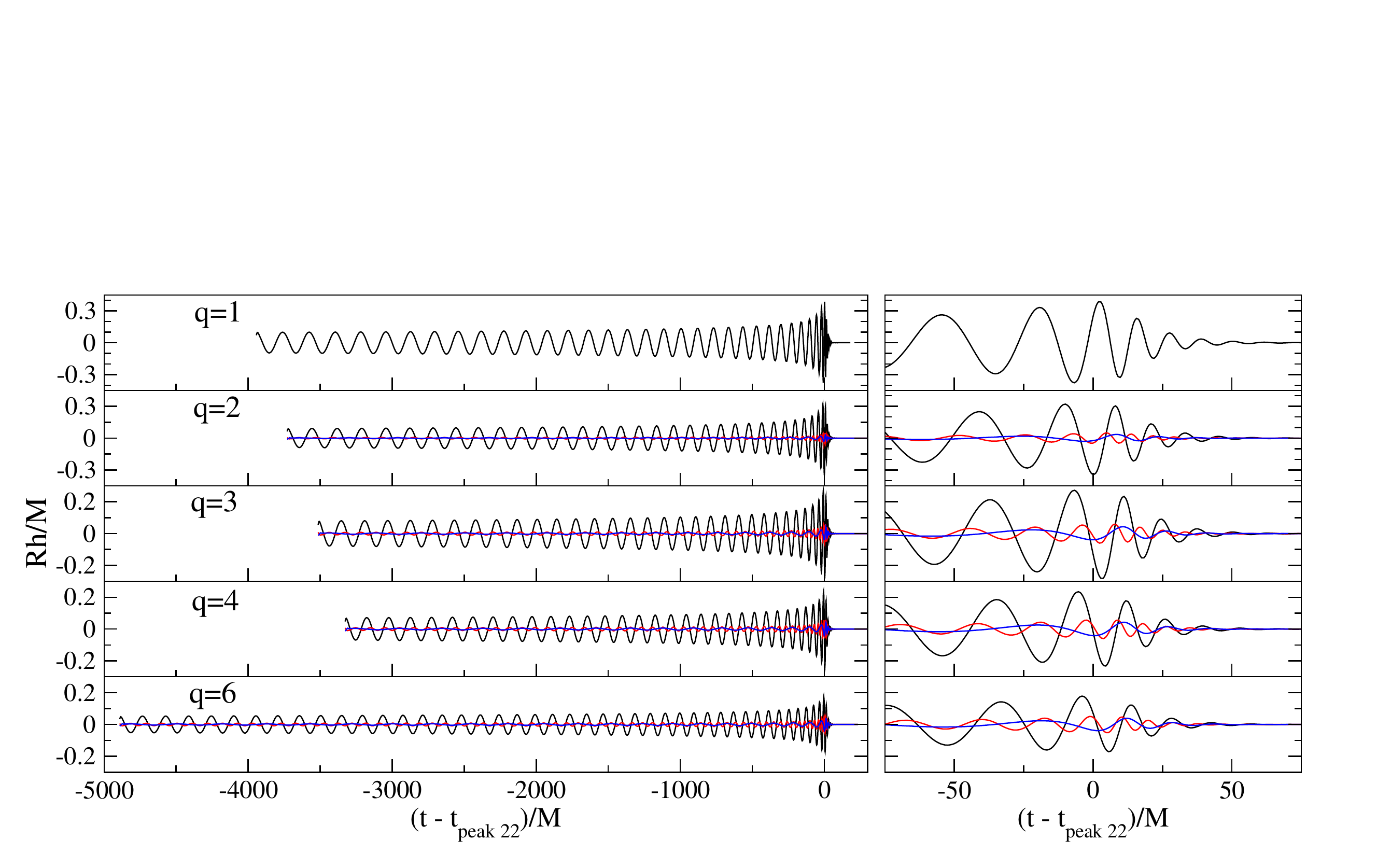}
\caption{\label{Fig:Waveforms} Gravitational waveforms for
  $q=1,2,3,4,6$ that have been extrapolated to infinity. The black
  curve is the $h_{22}$ mode, the red line is the $h_{33}$ mode, and
  the blue line is the $h_{21}$ mode. Only the real parts have been
  plotted.  The $x$-axis has been time-shifted so that $0$ indicates
  the merger, as determined by the peak of the extrapolated $h_{22}$
  mode waveform, for each mass ratio. The left-hand panels show the
  full coalescence: inspiral, merger, ringdown. The right-hand panels
  show a close-up of the merger and ringdown. Only the $h_{22}$ mode
  is shown for $q=1$, since the odd-m modes do not appear here.}
\end{figure*}

For unequal-mass BBHs, it is important to consider reflection
coefficients for higher-order modes, since the amplitude of these
modes relative to the dominant (2,2) mode increases with mass ratio
(see Fig.~\ref{fig:HigherModesa}).  For example, the reflection
coefficients for both the (2,1) mode and the (2,2) mode are given by
Eq.~(\ref{eq:sigma2}), but the (2,1) mode has twice the wavelength of
the (2,2) mode, reducing $k R_{\rm Bdry}$ by a corresponding factor of
2.  Consequently the reflection coefficient $\sigma_{21}$ of the (2,1)
mode is a factor $2^4=16$ times larger than the reflection coefficient
$\sigma_{22}$ of the (2,2) mode.  If we assume that the impact on the
phase error is proportional to the amplitude of the reflected waves,
then the relative importance of reflections of the (2,1) mode and the
(2,2) mode is given by the ratio
\begin{equation}\label{eq:ReflectionRatio}
Q_{m=1,m=2} \equiv \frac{A_{21} \sigma_{21}}{A_{22} \sigma_{22}},
\end{equation}
where $A_{2 1}$ and $A_{2 2}$ are the amplitudes of the (2,1) and
(2,2) modes, respectively.  Note that in the limit of large radii,
$Q_{m=1,m=2}$ is independent of boundary radius (because $R_{\rm
  Bdry}$ cancels out of the ratio $\sigma_{21}/\sigma_{22}$) and
independent of GW extraction radius (because the extraction radius
cancels out of the ratio $A_{21}/A_{22}$).  Looking up the amplitudes
of the (2,1) and (2,2) modes from Fig.~\ref{fig:HigherModesa}, and
using $\sigma_{21}/\sigma_{22}=16$ results in the numerical
values shown in Table~\ref{Table:ReflCoeff} (note that for these
calculations, the amplitudes were taken at a specific time during the
inspiral when they are still fairly constant). From this table, we
conclude that with our semi-absorbing (constraint preserving plus
freezing-$\Psi_0$) boundary conditions, the impact of the (2,1)
reflections on the overall phase error is comparable to that of the
(2,2) reflections, especially as the mass ratio increases to $q=4$ or
higher. With boundary conditions that are less than semi-absorbing,
the error contributions would be even higher.

%%%%%%%%%%%%%%%%%%%%%%%%%%%%%%%%%%%%%%%%%%%%%%%%%%%%%%%%%%%%%%%%
\subsection{Properties of gravitational radiation}
\label{sec:GW}
%%%%%%%%%%%%%%%%%%%%%%%%%%%%%%%%%%%%%%%%%%%%%%%%%%%%%%%%%%%%%%%%

Fig.~\ref{Fig:Waveforms} shows the waveforms for our 15-orbit
inspiral, merger and ringdown, as measured by $(R/M)h_{\ell m}$. All
these waves have been extrapolated to infinity. We show the top three
modes: $(2,2), (3,3), (2,1)$. Notice that the amplitude of the $(2,2)$
mode decreases as the mass ratio increases, but the amplitudes of the
other modes stay approximately the same. Further notice that the
wavelength of the $(2,1)$ mode is about twice that of the $(2,2)$
mode. This is a general property: for a given $\ell$, the wavelength
of the waveform is typically proportional to $1/|m|$.

\begin{figure}
\includegraphics[width=0.95\columnwidth]{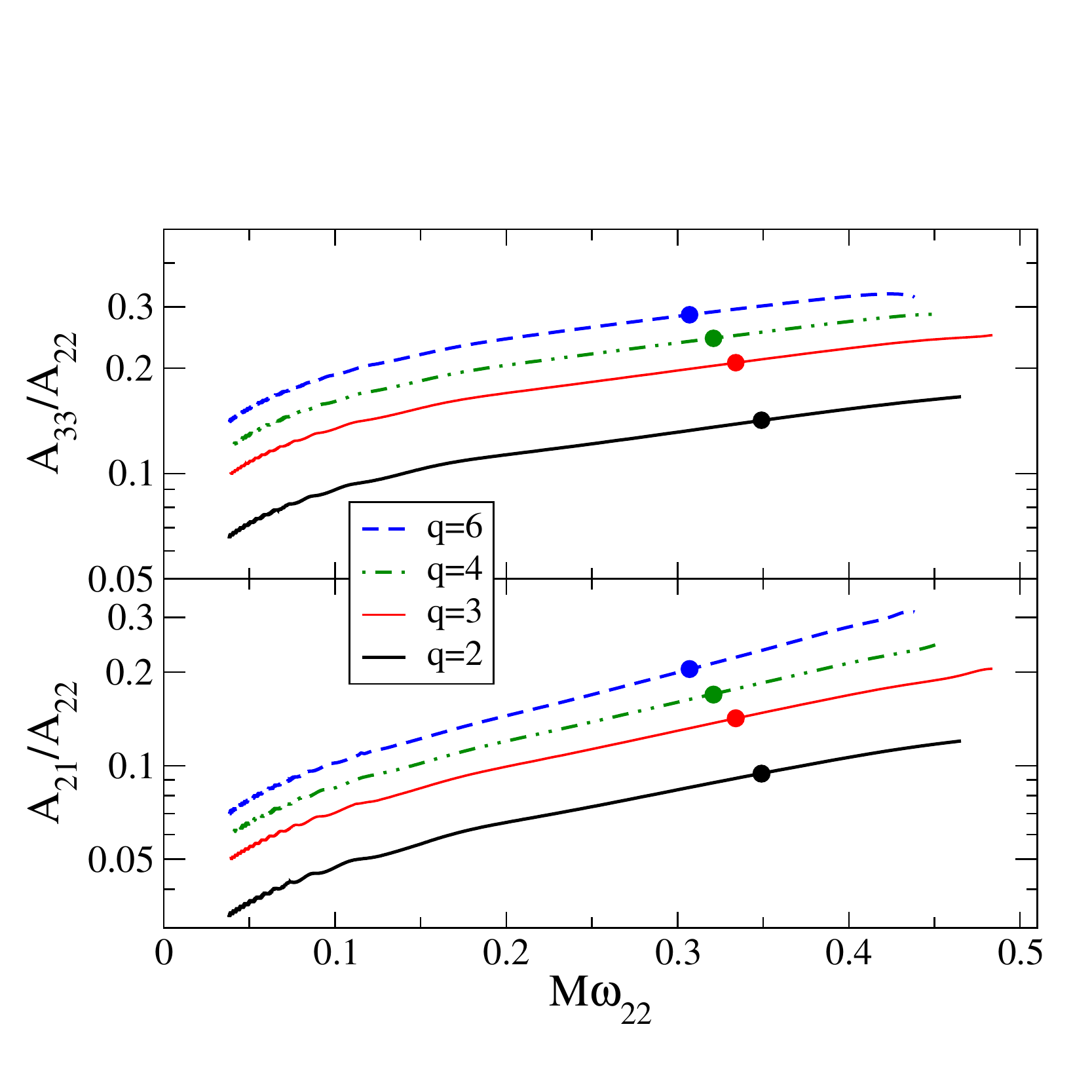}
\caption{\label{fig:HigherModesa} Amplitudes of $h_{33}$ (top) and
  $h_{21}$ (bottom) modes, normalized by the amplitude of the leading
  $h_{22}$ mode, for mass ratios $q=2,3,4,6$. Relative amplitudes are
  plotted versus the frequency of the $h_{22}$ mode. The filled
  circles indicate the frequencies where the amplitude of the $h_{22}$
  mode peaks.}
\end{figure}

The relative importance of the $(3,3)$ and $(2,1)$ mode amplitudes to
that of the $(2,2)$ mode is shown for the inspiral and merger in
Fig.~\ref{fig:HigherModesa} (top panel: $(3,3)$ mode, bottom panel:
$(2,1)$ mode). This figure clearly shows that the higher order modes
grow in relative significance as the mass ratio increases. At frequency
$M\omega_{22}=0.06$, the ratio $A_{33}/A_{22}$ ranges from $0.08$ (for
$q=2$) to $0.16$ (for $q=6$), and $A_{21}/A_{22}$ from $0.04$ (for
$q=2$) to $0.08$ (for $q=6$).  At the peak of the $h_{22}$ waveform
(indicated by the filled circles in Fig.~\ref{fig:HigherModesa}),
$A_{33}/A_{22}=0.14$ for $q=2$ and $0.28$ for $q=6$;
$A_{21}/A_{22}=0.09$ for $q=2$ and $0.20$ for $q=6$.

%%%%%%%%%%%%%%%%%%%%%%%%%%%%%%%%%%%%%%%%%%%%%%%%%%%%%%%%%%%%%%%%
\subsection{Black hole Spin \& Tidal spin-up}
%%%%%%%%%%%%%%%%%%%%%%%%%%%%%%%%%%%%%%%%%%%%%%%%%%%%%%%%%%%%%%%%

\begin{figure*}
\includegraphics[scale=0.5]{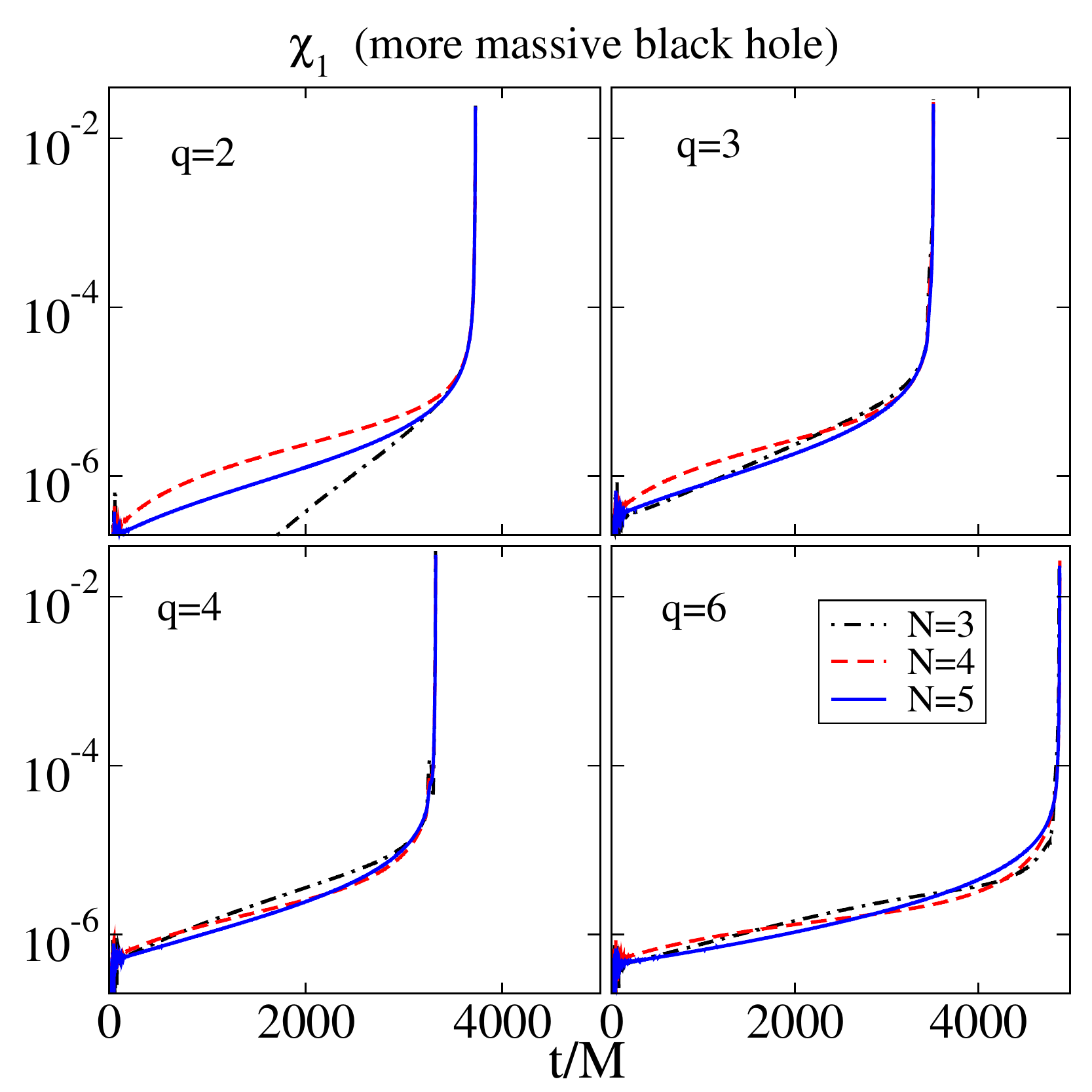}
$\qquad$
\includegraphics[scale=0.5]{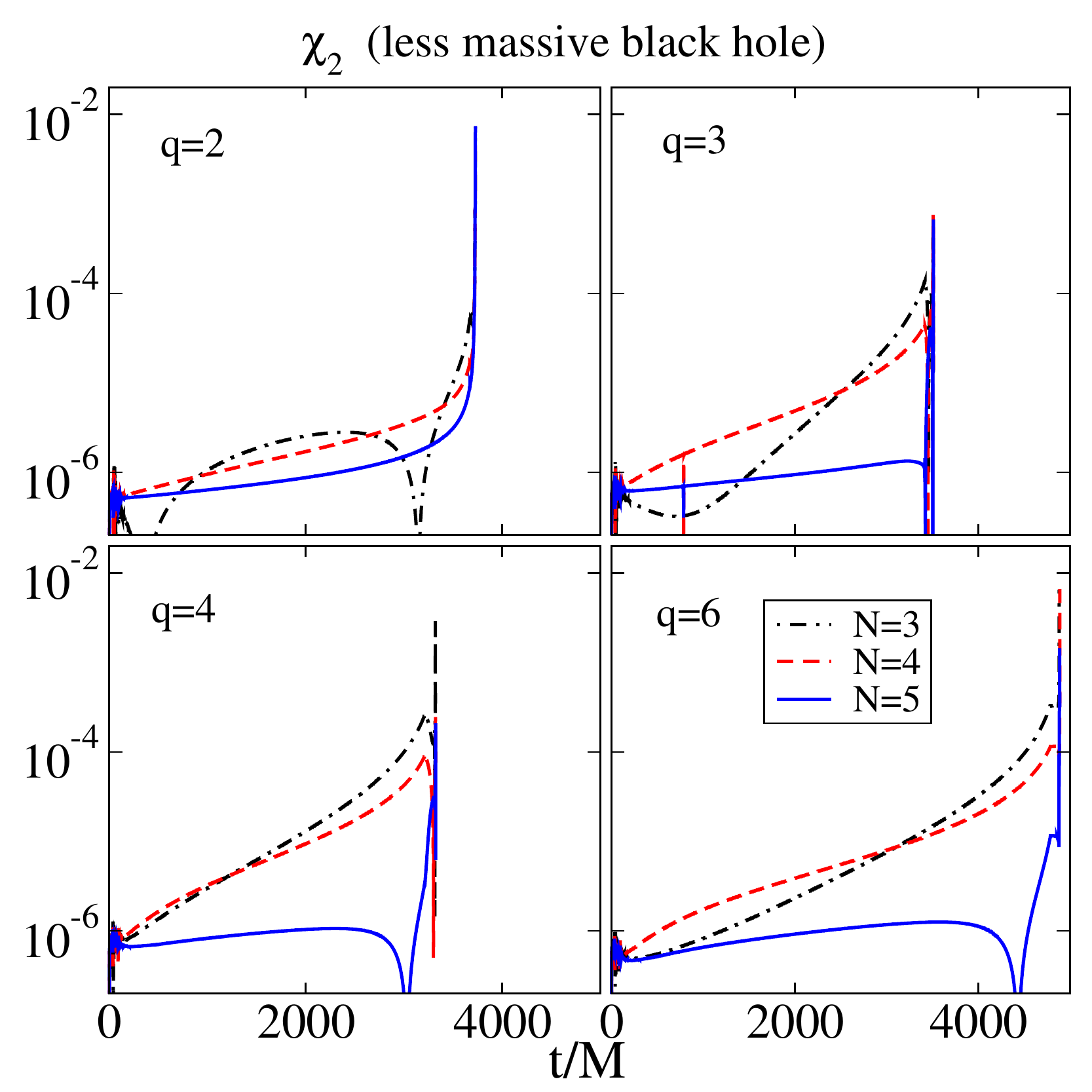}
\caption{\label{Fig:DimlessSpinHoleA} Convergence test of the
  dimensionless black hole spins $\chi=S/M^2$.  The left panel shows
  data for the more massive black hole, the right panel for the less
  massive black hole.  For each mass ratio, three resolutions are
  shown, labeled (N=3,4,5).  The spin of the
  more massive black hole, $\chi_1$ is convergently resolved and is
  monotonically growing during the simulation.  The spin of the
  smaller black hole $\chi_2$, is consistent with $\chi_2=0$ within
  numerical errors. }
\end{figure*}

We measure black hole spins by a surface integral on the apparent
horizon that utilizes approximate Killing vectors computed from a
minimization principle~\cite{Lovelace2008}.  We denote the
dimensionless spin by $\chi_A\!=\!S_A/M_A^2$ where $A\!=\!1$ indicates the
more massive black hole, and $A\!=\!2$ the less massive one.  At $t=0$,
both black hole spins are very small: $\chi_i(t=0)<10^{-8}$. This is
expected since $\chi_A=0$ is enforced as part of the initial data
construction, cf. Sec.~\ref{sec:ID}.  During the initial relaxation of
the initial data, the black hole spins increase to a few parts in
$10^{-7}$.  Subsequently, $\chi_1$ slowly increases during the
inspiral (with spin rotation axis parallel to the orbital angular
momentum).  This increase is convergently resolved, as shown in the
left panel of Fig.~\ref{Fig:DimlessSpinHoleA}.  In contrast, the spin
of the smaller black hole $\chi_2$ remains closer to zero, as shown in
the right panel if Fig.~\ref{Fig:DimlessSpinHoleA}.  For mass ratios
$q=3,4,6$, $\chi_2$ is consistent with zero within truncation error.
For $q=2$, there is a marginal detection of non-zero spin at late
times $t\gtrsim 3000M$.

\begin{figure}
\includegraphics[scale=0.5]{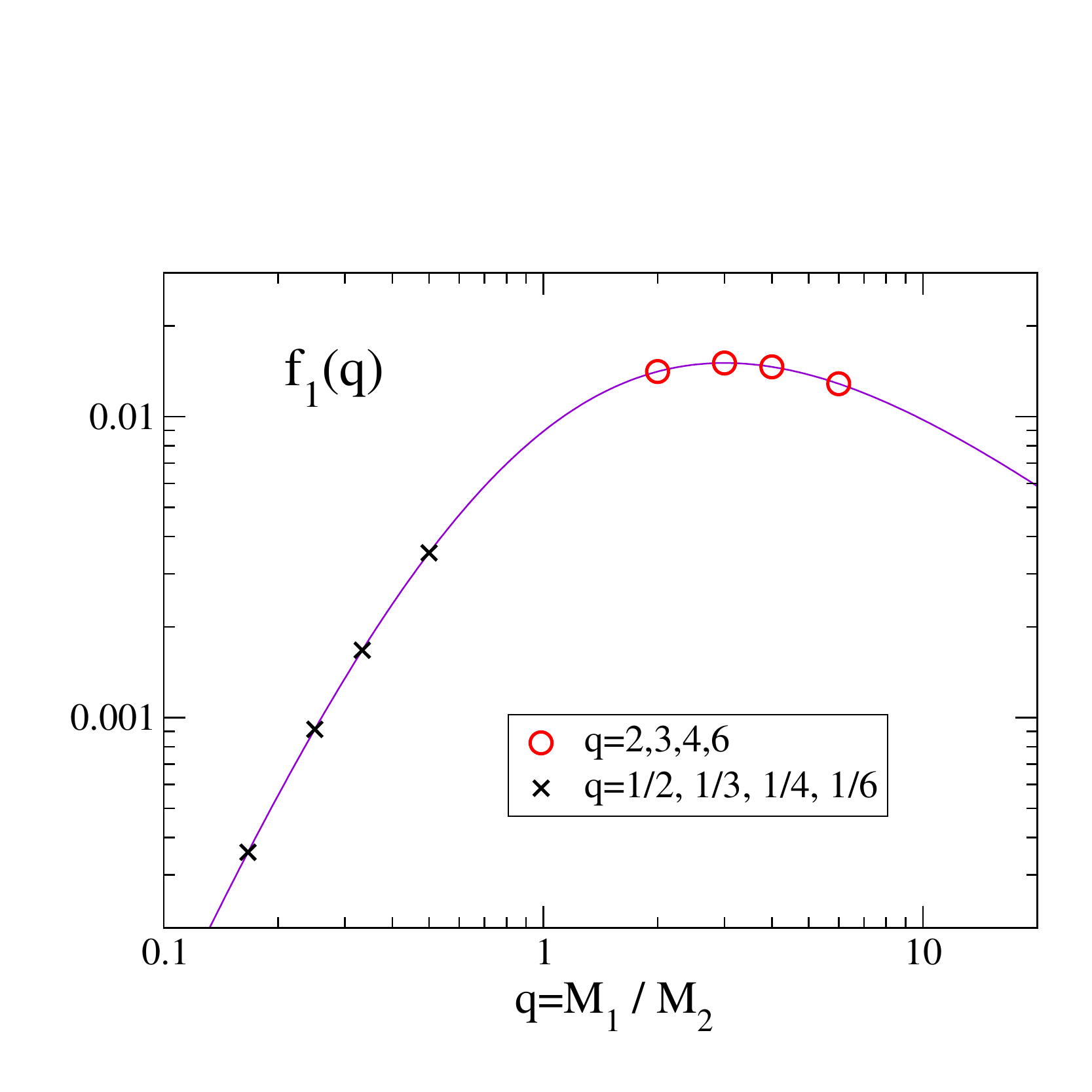}
\caption{\label{fig:f}
The coupling coefficient $f$ that determines the magnitude of the
  change of the spin $\chi_1$ during the inspiral as a function of the
  mass ratio $q=M_1/M_2$.  The red circles denote the coefficients for
the large black hole for the mass ratios simulated here.  The crosses
denote the coefficient for the small black hole, which can be obtained 
from the same plot at the inverse mass ratio.}
\end{figure}

We interpret the monotonically increasing spin $\chi_1$ as evidence of
tidal spin-up of non-rotating black holes.  To investigate this
process in more detail, we consider the spin $\chi_1$ as
a function of the orbital frequency.  Alvi~\cite{Alvi:2001mx} derived
tidal spin-up as a function of binary coordinate separation $b/M$.
Converting his formula into a function of the 
orbital frequency 
(which heuristically should be less gauge dependent)
via $M/b=(M\Omega)^{2/3}$, one obtains 
\begin{align}
\chi_1-\chi_{1,\infty}=&\frac{\eta M_1}{4M}(1+3\chi_{1,\infty}^2)\\
\nonumber
&\times \left(-\frac{\chi_{1,\infty}}{4}\, (M\Omega)^{4/3} 
+\frac{2r_{1,\infty}}{7M}\,(M\Omega)^{7/3}\right).
\end{align}

$\chi_{1,\infty}$ is the spin magnitude of black hole 1 at infinite separation, 
and $r_{1,\infty}=M_1(1+\sqrt{1-\chi_{1,\infty}^2})$
is the corresponding horizon radius.  Dropping terms quadratic in 
$\chi_{1,\infty}$ because of their small size, this equation simplifies to
\begin{equation}
\chi_1=\chi_{1,\infty}\left(1-\frac{\eta M_1}{16 M}(m\Omega)^{4/3}\right)
+\frac{\eta M_1^2}{7M^2}(m\Omega)^{7/3}.
\end{equation}
Furthermore, the expression in parentheses in the first term on the
right hand side is so close to unity that the deviation from unity is
irrelevant given the small value of $\chi_{1,\infty}$.  Approximating
this parenthesis by unity, we finally find
\begin{equation}\label{eq:chiAlvi}
\chi_1 = \chi_{1,\infty} + f_1\; (M\Omega)^{7/3}
\end{equation}
with the coefficient
\begin{equation}
f_1 = \frac{\eta M_1^2}{7M^2} = \frac{q^3}{7(1+q)^4}.
\end{equation}
Therefore, we see that the spin $\chi_1(M\Omega)$ should follow a
power law in frequency $M\Omega$.

The magnitude of the change in the spin is determined by the
coefficient $f_1(q)$, which is plotted in Fig.~\ref{fig:f}.  The red
circles denote the values of this coefficient for the large black hole
in our simulations: The mass ratios considered here all result in
almost maximal tidal coupling, for maximal spin-up of the large black
hole.  In contrast, the black crosses denote the spin coupling
coefficient for the small black hole.  The spin coupling coefficient
for the small black hole is smaller by a factor between 4 ($q=2$) and
36 ($q=6$), indicating that the smaller black hole will be much less
susceptible to tidal spin-up.  Therefore, from the perturbative
analysis of tidal coupling, we expect that the larger black hole in
all our simulations will be spun up by approximately similar amounts,
and that the small black hole will be spun up significantly less.
This expectation is already borne out in
Fig.~\ref{Fig:DimlessSpinHoleA}, where we were able to resolve the
spin-up of BH 1, but not the (smaller) spin-up of BH 2.

\begin{figure}
\includegraphics[scale=0.5]{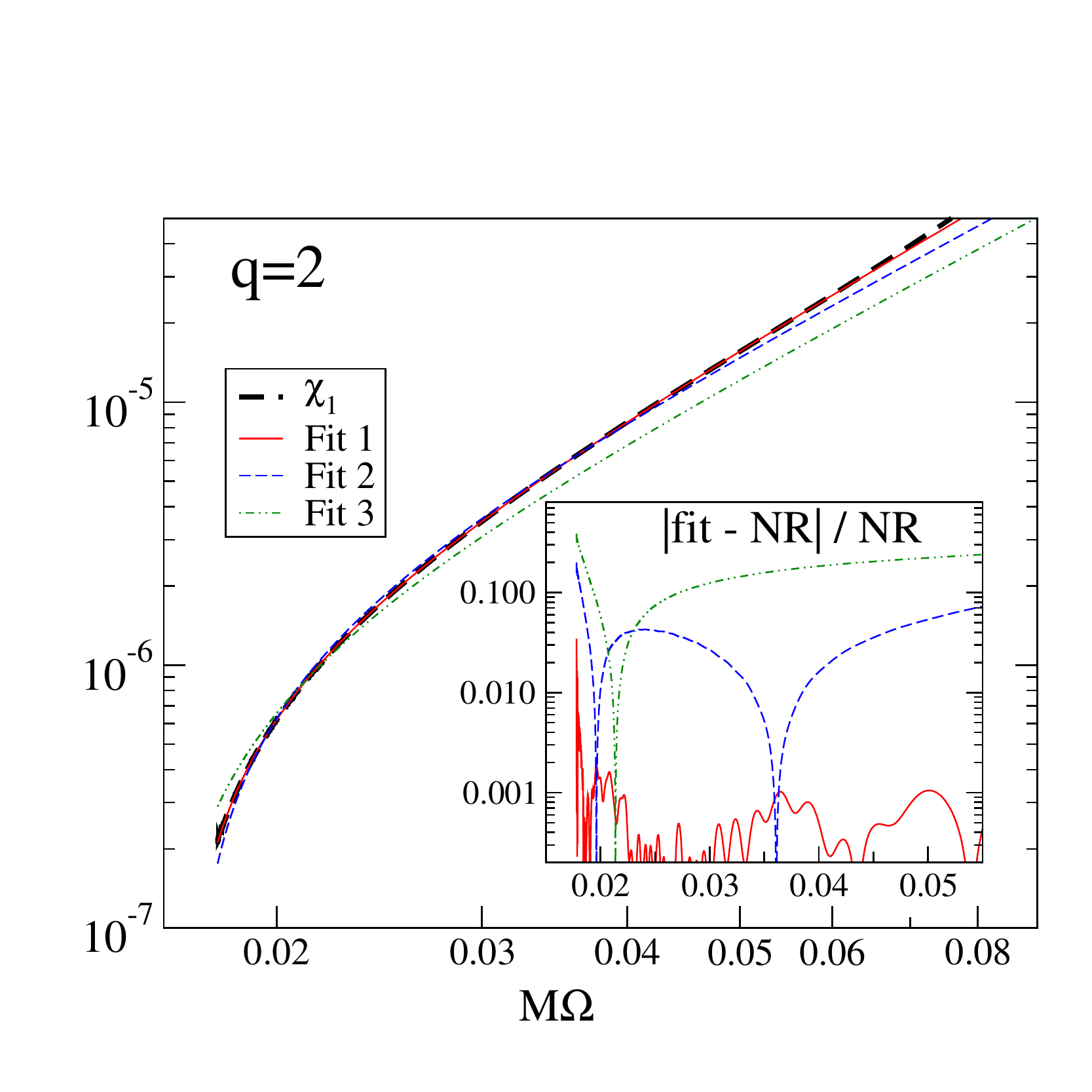}\\[1em]
\includegraphics[scale=0.5]{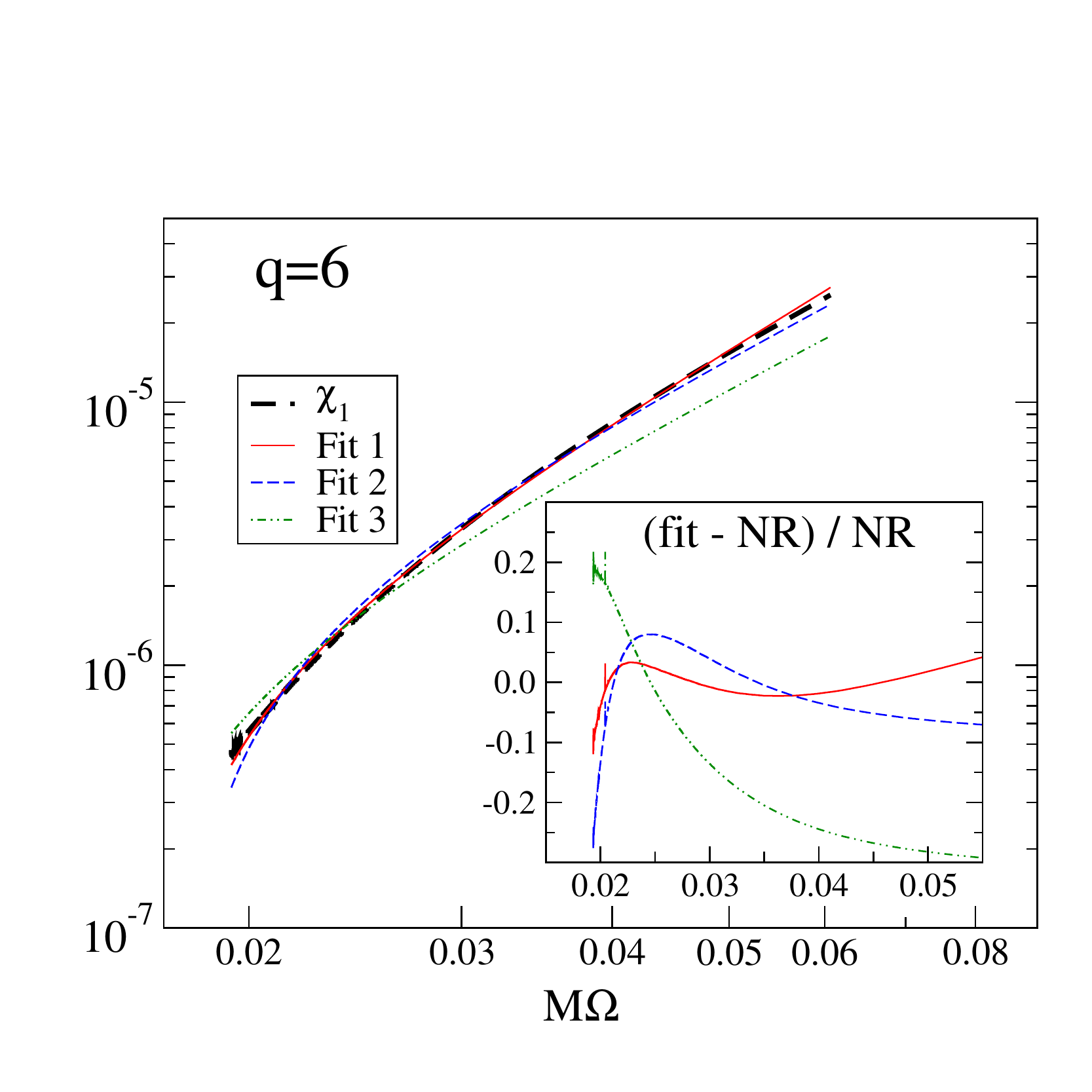}
\caption{\label{fig:SpinFits}Dimensionless spin $\chi_1$ of the larger
  black hole as a function of the orbital frequency $M\Omega$.
  Plotted is the numerical data, and three fits to the data, fitted in
  the interval $M\Omega\le 0.055$.}
\end{figure}

\begin{table}[b]
\begin{tabular}{c|ccc|cc|c}
 & \multicolumn{3}{|c|}{Fit 1} 
  & \multicolumn{2}{|c|}{Fit 2} &
  \multicolumn{1}{c}{Fit 3} \\
$q\;\;$ & \multicolumn{3}{|c|}{$A_0 + A_1 (M\Omega)^{A_2}$} 
  & \multicolumn{2}{|c|}{$B_0 + B_1 f_1 (M\Omega)^{7/3}$} &
  \multicolumn{1}{c}{$C_0+f_1(M\Omega)^{7/3}$} \\
& $10^6A_0$ & $A_1$ & $A_2$ & $\quad 10^6B_0\quad$ & $\quad B_1\quad$  &  $10^6C_0$  \\ \hline
2 &  -0.95 & 0.0362 & 2.57 &   -1.26 & 1.23 &    -0.88 \\
3 &  -1.10 & 0.0496 & 2.64 &   -1.59 & 1.29 &    -0.99 \\
4 &  -1.25 & 0.0474 & 2.62 &   -1.78 & 1.34 &    -0.96 \\
6 &  -0.81 & 0.0602 & 2.74 &   -1.39 & 1.34 &    -0.74 \\
\end{tabular}
\caption{\label{tab:SpinFits} Fitting parameters for fits to the $\chi_1(M\Omega)$ data.}
\end{table}

Fitting the numerical data $\chi_1(M\Omega)$ to the functional form of
Eq.~(\ref{eq:chiAlvi}) with the one free fitting parameter
$\chi_{1,\infty}$ results in a moderately good fit. The fit can be
improved if the coefficient $f_1$ is also fitted for, and can be
improved further by also allowing the exponent to vary, i.e. a
power-law fit with an offset.  The results of these fits (which we
refer to as Fit 3, Fit 2, and Fit 1, respectively), are shown in
Table~\ref{tab:SpinFits}.  Figure~\ref{fig:SpinFits} plots the fits
and their residuals for mass ratios $q=2$ and $q=6$.  All fits were
performed over the numerical data up to orbital frequency
$M\Omega=0.055$\footnote{Beyond this frequency, we modify the gauge in
  the simulation, which leads to artifacts in $\chi_1(\Omega M)$.}.
As can be seen from the insets of Fig.~\ref{fig:SpinFits}, the more
general Fit 1 is superior to a fit with fixed exponent 7/3 (Fit 2),
which in turn is superior to the one-parameter Fit 3 of
Eq.~(\ref{eq:chiAlvi}).  For $q=2$, the residual of Fit 1 is almost
two orders of magnitude smaller than for fits 2 and 3.  Coefficient
$A_2$ in Table~\ref{tab:SpinFits} shows that the numerical data
prefers a power law with a slightly {\em larger} exponent of roughly
$8/3$ instead of the expected $7/3$.  If the exponent is fixed to
$7/3$, then coefficient $B_1$ indicates that the overall magnitude of
the spin-evolution is larger in the numerical simulation by about a
factor of $1.3$ relative to the expected behavior
Eq.~(\ref{eq:chiAlvi}).  All fits indicate fairly consistently that
the spin of the large black hole at infinite separation would be
around $10^{-6}$, anti-aligned with the orbital angular momentum
(cf. coefficients $A_0, B_0, C_0$).

These results are enticing and suggestive.  However, we caution the
reader that the observed effects are very small, with changes to the
dimensionless spin of order $10^{-5}$.  Before drawing firm
conclusions, one must establish that the numerical data is accurate
enough by performing a three-fold convergence test.  First, the
resolution of the numerical evolution must be varied to determine that
Einstein's equations are solved with sufficient accuracy. 
This we have done.  However, in addition to this numerical convergence
test, the resolution of the apparent
horizon finder must be varied to ascertain that the apparent horizon
is found with adequate accuracy.  And finally, the resolution of the
eigenvalue solver that computes the approximate helical Killing
vectors on the apparent horizon (cf. Appendix of \cite{Lovelace2008})
must be varied to check that the approximate Killing vectors are
calculated accurately enough.  Unfortunately, we did not output enough
data during the numerical evolutions to perform the second two
convergence tests.

In addition, further work would be needed to ascertain that the
approximate Killing vectors (and the spin computed using these,
cf.~\cite{Lovelace2008}) are indeed generating a spin compatible with
the spin definitions of the perturbative work~\cite{Alvi:2001mx}.
Because of all these cautionary comments, and insufficient numerical
data, we postpone quantitative results about tidal spin-up to future
work.

%%%%%%%%%%%%%%%%%%%%%%%%%%%%%%%%%%%%%%%%%%%%%%%%%%%%%%%%%%%%%%%%
\subsection{Remnant properties}
%%%%%%%%%%%%%%%%%%%%%%%%%%%%%%%%%%%%%%%%%%%%%%%%%%%%%%%%%%%%%%%%

\begin{figure}
\centerline{\includegraphics[scale=0.5]{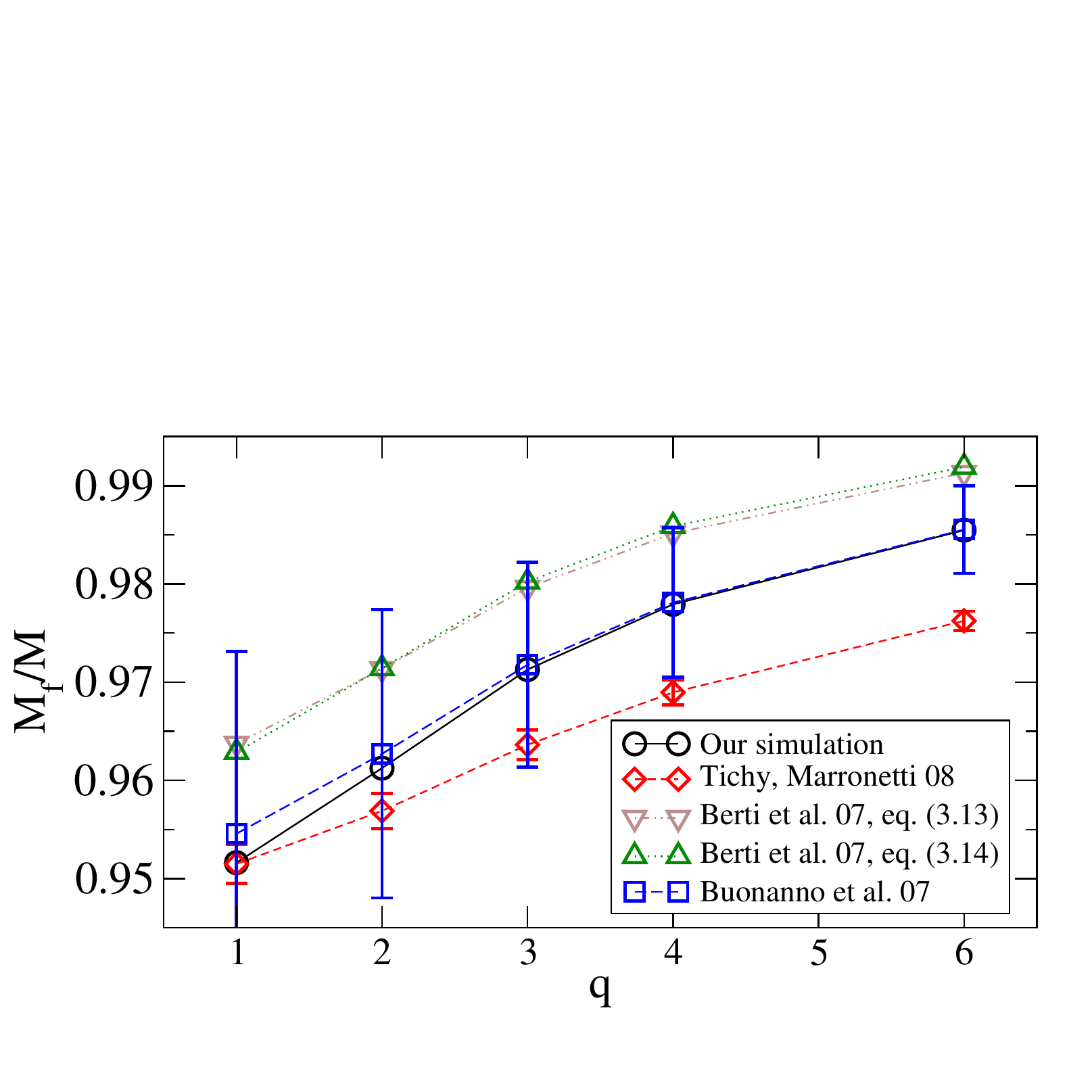}}
\caption{\label{fig:FinalMass}$M_f/M$ as function of $q$.  Also
  shown are the results from the fitting formula of Tichy and
  Marronetti~\cite{Tichy2008}, the analytical prediction of 
  Berti et al.~\cite{Berti2007},
  and the fit of Buonanno et al.~\cite{Buonanno2007} to numerical data.}
\end{figure}

\begin{figure}
\centerline{\includegraphics[scale=0.5]{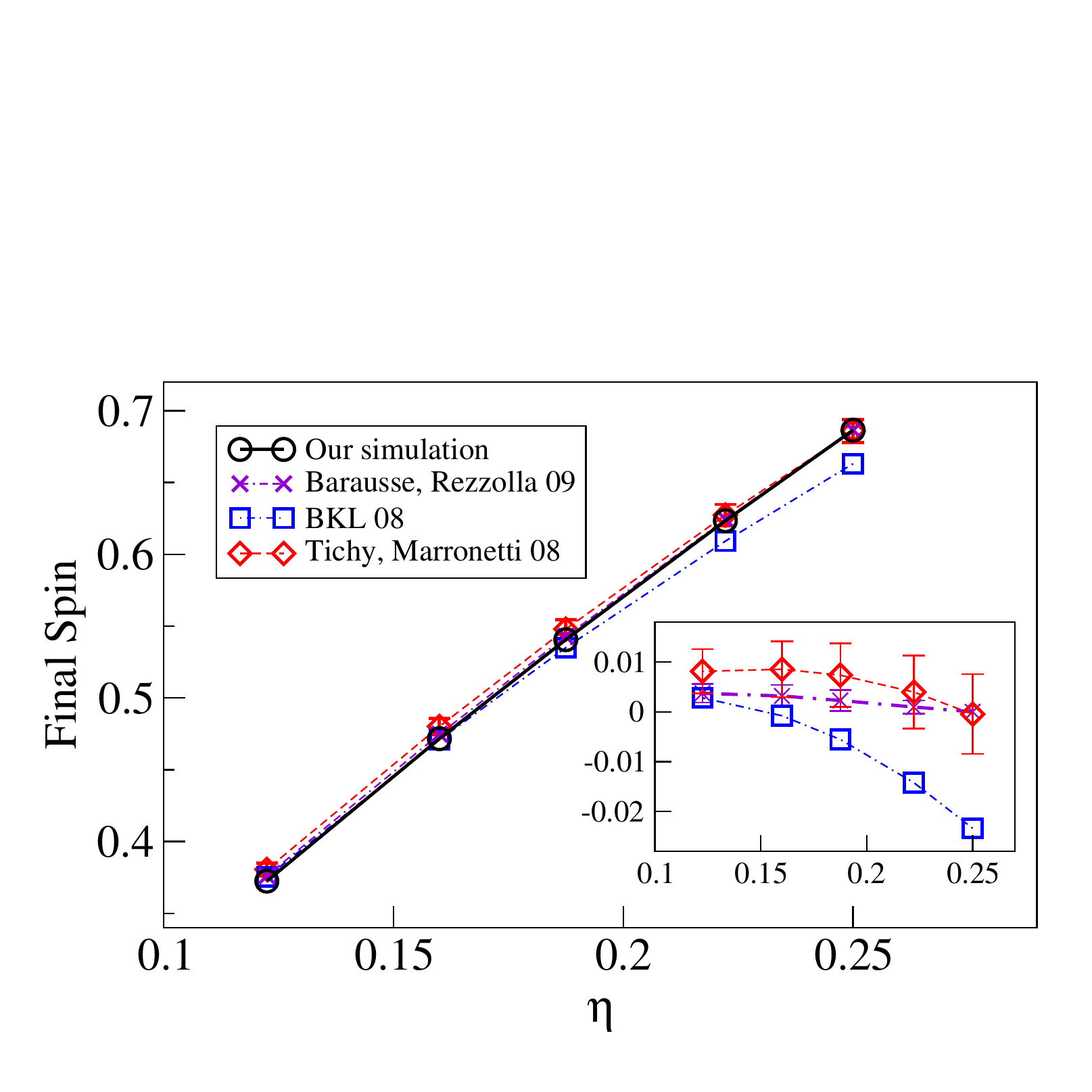}}
\caption{\label{fig:FinalSpin}$S_f/M_f^2$ as function of symmetric
  mass ratio $\eta$.  Also shown are the results of fitting formulas
  and estimates from Barausse and Rezzolla~\cite{Barausse2009}, from
  Buonanno, Kidder and Lehner (BKL)~\cite{Buonanno2008a}, and from
  Tichy and Marronetti~\cite{Tichy2008}. The inset shows the
    difference $\chi_{\rm fit}-\chi_{\rm NR}$ between fitting formula
    and our numerical results.}
\end{figure}

Figures~\ref{fig:FinalMass} and~\ref{fig:FinalSpin} show the mass and
spin of the remnant black hole (computed using approximate
Killing vectors on the apparent horizon~\cite{Lovelace2008,OwenThesis,
Dreyer2003,Cook2007}) as a function of mass ratio $q$. These
quantities are also listed in Table~\ref{Table:RunParameters}.
Several fitting formulas in the literature give good agreement with
the remnant spin and are plotted in
Figure~\ref{fig:FinalSpin}. Analytical predictions of the final mass
do not agree as quite as well, as seen in Figure~\ref{fig:FinalMass}; however,
the formula of Buonanno et al.~\cite{Buonanno2007}, which is a fit to numerical
relativity results, shows better agreement.

For unequal-mass binaries, linear momentum is carried off
anisotropically by gravitational waves, leading to a recoil of the
remnant black hole.  The recoil speed of the remnant can be computed
from the gravitational-wave momentum flux at infinity.  To do this, we
start with the Newman-Penrose quantity $\Psi_4$, extracted from our
simulations and extrapolated to infinite radius using the procedure of
Boyle and Mrou\'e~\cite{Boyle-Mroue:2008}.  The momentum flux depends
on the first time integral of $\Psi_4$, and computing this time
integral requires two integration constants, which we determine by the
procedure outlined in Appendix B of Ref.~\cite{Boyle:2008}. This
procedure involves a minimization over a time interval $[t_1,t_2]$,
where $t_1$ and $t_2$ can be chosen arbitrarily.  We find that varying
the integration-constant parameters $t_1$ and $t_2$ in the range $t_1
\in [1000M,1400M]$ and $t_2 \in [2600M,3000M]$ changes $v_{\rm kick}$
by only a tenth of a percent.  Once we have the time integral of
$\Psi_4$, we compute the gravitational-wave momentum flux by the
procedure of Ref.~\cite{RuizTakAlcNunez2007}, keeping all $Y_{\ell m}$
modes through $\ell=6$.  The time integral of the momentum flux gives
the total radiated 3-momentum $\vec{P}$, and the recoil velocity is
$\vec{v}\equiv -\vec{P}/M_f$.  Note that the recoil velocity can 
alternatively be computed
by a time derivative of the Regge-Wheeler-Zerilli strain $h_{\ell m}$ 
rather than a time integral of $\Psi_4$.  We use the latter method because
differentiation amplifies noise in the waveform to the extent that for the
runs shown here, the former method would require smoothing 
put in by hand.

\begin{figure}
\centerline{\includegraphics[scale=0.5]{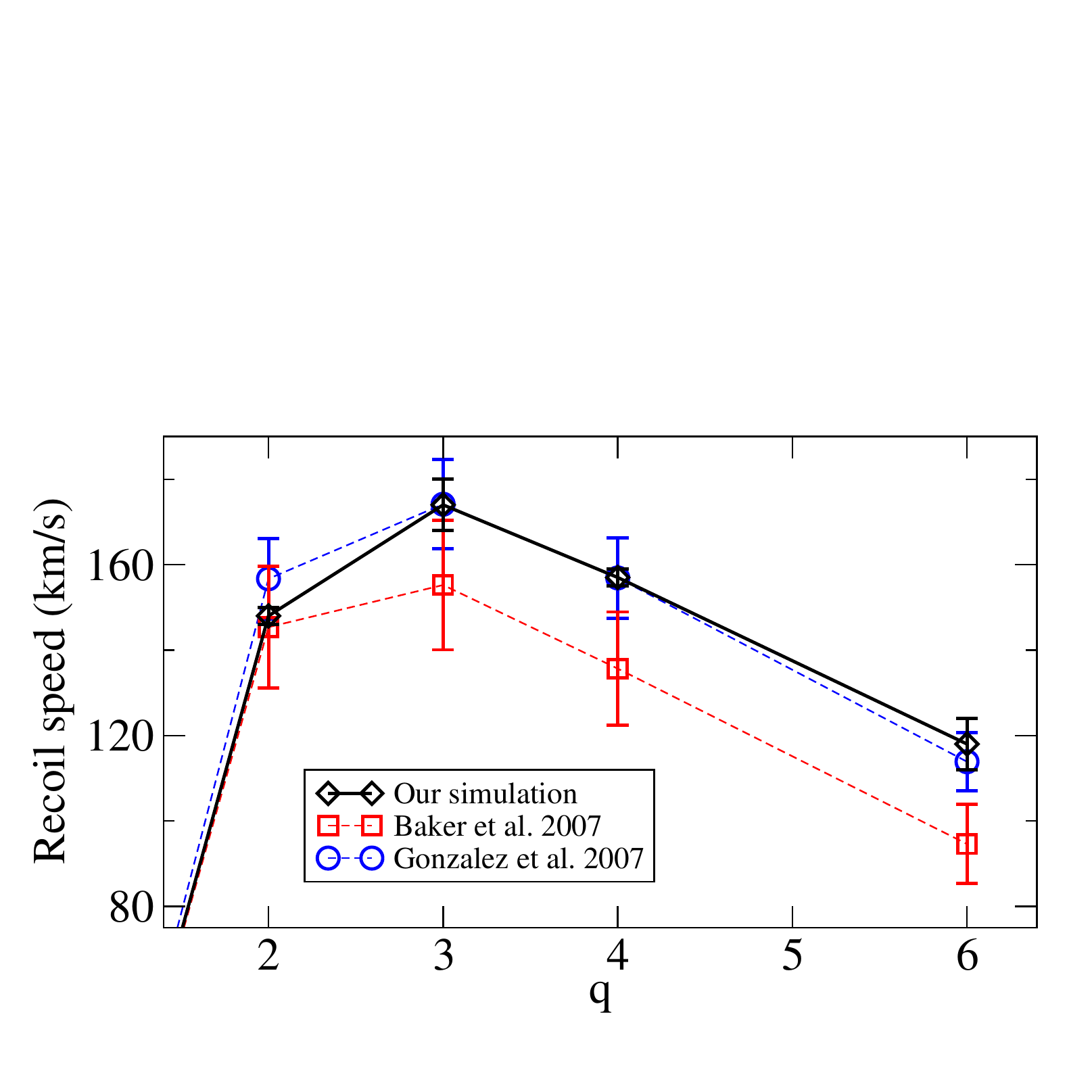}}
\caption{\label{fig:FinalKick}$|v_{\rm kick}|$ as function of $q$.
Also shown are the results of fitting formulas and estimates
from Baker {\it et al.}~\cite{Baker2007} and from Gonzalez {\it et 
al.}~\cite{Gonzalez2007}.}
\end{figure}

The recoil speed $v_{\rm kick}\equiv|\vec{v}|$ of the remnant is listed in
the last column of Table~\ref{Table:RunParameters}.  We estimate
several sources of uncertainty, which are listed in
Table~\ref{Table:KickErrors}.  Numerical truncation error is estimated
by taking the difference of $v_{\rm kick}$ computed using the highest
and second-highest numerical resolutions; this is the dominant source
of error for two of our simulations.  The uncertainty in extrapolating
the waveform to infinity is estimated by comparing $v_{\rm kick}$
computed using waves extrapolated using 3rd order
polynomials~\cite{Boyle-Mroue:2008} versus an identical calculation
using 4th order polynomials.  The error associated with truncating
$Y_{\ell m}$ modes for $\ell>6$ in the momentum flux is estimated by
comparing with an identical calculation where we retain only $\ell\leq
5$.  Initial data effects such as the initial pulse of junk radiation
add a spurious recoil of about 1 to 2 km/s, depending on the run.
There is an additional small error that results from neglecting the
recoil that occurs in the early inspiral between $t=-\infty$ and the
start of our simulations; this neglected recoil can be estimated to
2PN order using Eq. 22 of ref.~\cite{Blanchet2005b}, which yields
about 0.5 km/s for the cases shown here.  Figure~\ref{fig:FinalKick}
plots the recoil versus mass ratio for our simulations and for two
fitting formulas in the literature.  We find good agreement.

\begin{table}[b]
\begin{tabular}{|c|cccccc|} 
  \hline 
  $\;q\;$ & $v_{\rm kick}$        & 
  $\delta v_{\rm kick}^{\rm T}$    &
  $\delta v_{\rm kick}^{\rm E}$    &
  $\delta v_{\rm kick}^{Y_{\ell m}}$ &
  $\delta v_{\rm kick}^{\rm ID}$   &
  $\delta v_{\rm kick}^{t\to\-\infty}$ \\
  \hline
  2 & 148 & 0.7 & 0.4  & 1   & 1 & 0.4\\
  3 & 174 &  6  & 0.4  & 0.2 & 2 & 0.6\\
  4 & 157 & 1.2 & 0.4  & 0.3 & 2 & 0.6\\
  6 & 118 & 4.5 & 1    & 3   & 1 & 0.4\\
  \hline
\end{tabular}
\caption{\label{Table:KickErrors} Recoil velocity and uncertainties
  in km/s.  Uncertainties (left to right) are numerical truncation error,
  error in extrapolating waveforms to infinity, the effect of 
  using only a finite
  number of $Y_{\ell m}$ modes to compute the momentum flux, error involving
  initial transients (e.g. junk radiation), and the estimated recoil accumulated
  from $t=-\infty$ to the start of our simulation.
}
\end{table}

%#########################
\section{Discussion}
\label{sec:Discussion}
%#########################

This paper accomplishes several tasks with regard to simulations of
BBH systems.  Section~\ref{sec:ID} introduces an efficient formalism
to perform root-finding necessary to achieve desired initial data
parameters (masses, spins, center-of-mass frame).  Each function
evaluation during root-finding is an entire (expensive) initial-data
solve, so it is imperative to be able to perform this procedure with
as few function evaluations as possible.  The procedure introduced
here, based on approximate Newton-Raphson iteration, performs very
well.  As Fig.~\ref{fig:BBH_ID_Convergence} shows, one or two
high-resolution initial data runs are sufficient.  Since the
high-resolution solutions dominate the overall CPU cost, root-finding
can thus be accomplished with marginal extra cost.  This procedure has
since then been extended to superposed Kerr-Schild
data~\cite{Lovelace2008}.

We then give technical details about how to simulate unequal-mass
binaries with multi-domain spectral methods. In particular, we extend
the dual-frame formalism and control systems to unequal masses,
introduce eccentricity removal for unequal-mass binaries, and describe
algorithmic modifications performed during merger and ringdown.

The largest part of this paper documents a new series of unequal-mass,
non-spinning BBH simulations with mass ratios $q=2,3,4$ and $6$,
lasting between 15 and 22 orbits before merger.  We show that these
simulations have high accuracy, comparable to that of the equal-mass
simulation presented in~\cite{Boyle2007,Scheel2009}.  The total mass
is conserved during the inspiral to a few parts in $10^6$
(cf. Fig.~\ref{Fig:MassRelxn}), a convergence test on the (not
time-shifted gravitational wave phase) indicates that errors in our
second highest resolution run are a few tenths of a radian.  Given how
much more challenging a mass-ratio 6 simulation is, we are very
encouraged that the errors are only larger by a factor of $4$ relative
to the equal-mass simulation, cf. Fig.~\ref{Fig:GWPhaseConv22}.  By
moving the outer boundary, we establish furthermore, that effects due
to the outer boundary arise at the smaller level of $\sim 0.01$rad in
the waveform, as shown in Fig.~\ref{Fig:OuterBdryTests}.  We also
perform a convergence study on the subdominant (3,3) and (2,1) modes
of the gravitational radiation.  These subdominant modes become more
important with higher mass ratio (see \cite{Berti2007,HannamEtAl:2010}
and Fig.~\ref{fig:HigherModesa}), and we argue that this increases the
need for reflection minimizing boundary conditions, as those applied
here.  The final waveforms, extrapolated to infinite extraction
radius, are shown in Fig.~\ref{Fig:Waveforms}.

We then consider carefully the change in the spin of the larger black
hole. This change is broadly consistent with perturbative calculations
of black holes: The power law of the spin vs. orbital frequency is
rather well matched ($\sim 2.66$ vs. $7/3$), and the amplitude of the
change is also reasonably close, being off by a factor $\sim 1.3$.  A
more detailed comparison must, however, await more complete
convergence data, to allow comprehensive quantification of the error
in the numerical spin.  But nevertheless, these data point to the fact
that our simulations are in fact for a BBH where the larger black hole
started at infinite separation with a spin of $\sim 10^{-6}$
\emph{anti-aligned} to the orbital momentum.  Tidal spin-up increases
this spin during the early (not modeled) inspiral, so that the spin
passes through zero when our simulations commence.

Finally, we compare remnant properties and kick velocities. These are
found to be in reasonable agreement to various fitting formulae in the
literature.

An important result of this work is the accurate calculation of long
{\em subdominant} mode waveforms. These are needed for parameter
estimation, calculating physical quantities such as the gravitational
recoil, and for modeling analytic and phenomenological waveforms
(see~\cite{HannamEtAl:2010} and references therein). Furthermore,
recent results indicate that they are important for LIGO event
detection: Brown, Kumar and Nitz (in prep 2012) have found that for
$q>1.8$, the top subdominant modes must be taken into account in order
to achieve the usual signal to noise ratio loss criterion ``overlap
greater than 0.965''.  Pertinent factors used in these simulations
which have contributed to the achieved accuracy are: (i) our use of
semi-absorbing boundary conditions combined with the location of the
outer boundary, (ii) extrapolation to infinity, (iii) good numerical
resolution because of the length scale problem (which becomes more
severe for the subdominant modes), and (iv) pseudo-spectral
methods. In sum, we have been able to perform the first long and
accurate numerical simulations of unequal non-spinning binary black
holes with mass ratios as high as 6, with excellent convergence and
modest computational cost, even for the subdominant modes.

\begin{acknowledgments}
  We would like to thank Enrico Barausse, Duncan Brown and Abdul
  Mrou\'{e} for useful discussions, and Michael Boyle and Fan Zhang
  for performing wave extrapolations. We are also very grateful to
    the referee for his or her thorough reading and thoughtful
    suggestions. This work was supported in part by grants from the
  Sherman Fairchild Foundation to Caltech and Cornell and by NSF
  grants PHY-1068881 and PHY-1005655 and NASA grant NNX09AF97G at
  Caltech.  H.P. gratefully acknowledges support from the NSERC of
  Canada, from the Canada Research Chairs Program, and from the
  Canadian Institute for Advanced Research.  Computations were
  performed on the Syracuse University Gravitation and Relativity
  (SUGAR) Cluster, which is supported by Syracuse University and NSF
  award PHY-0600953; on the NSF XSEDE network under grant
  TG-PHY990007N; on the Zwicky cluster at Caltech, which is supported
  by the Sherman Fairchild Foundation and by NSF award PHY-0960291;
  and on the GPC supercomputer at the SciNet HPC
  Consortium~\cite{scinet}. SciNet is funded by: the Canada Foundation
  for Innovation under the auspices of Compute Canada; the Government
  of Ontario; Ontario Research Fund--Research Excellence; and the
  University of Toronto.
\end{acknowledgments}

\appendix*{}
\section{Non-overlapping spectral grid}
\begin{figure}
\centerline{\includegraphics[scale=0.5]{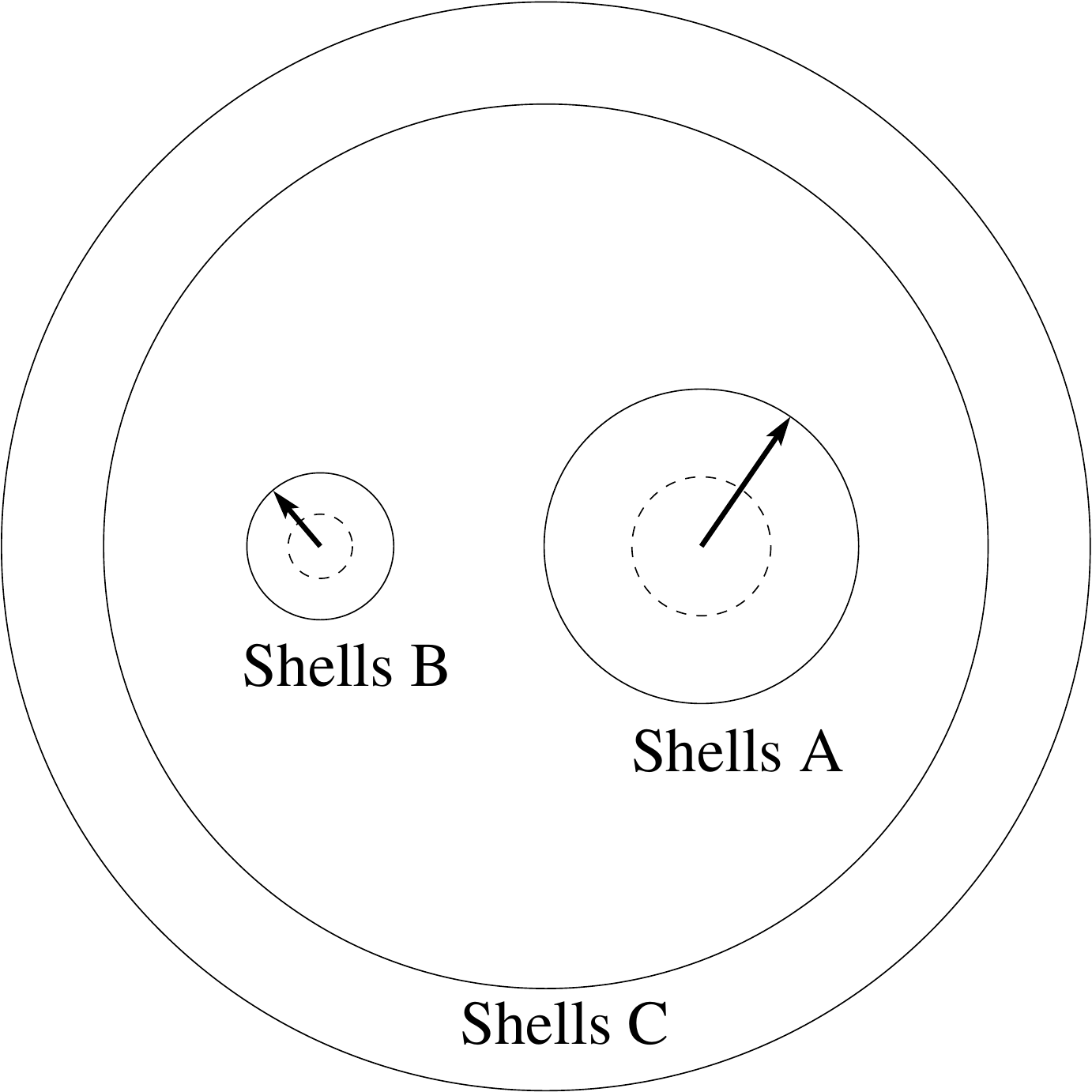}}
\caption{\label{fig:CutSphereShells} Schematic geometry of
  the spherical regions of the grid geometry. The outer radii
  of the regions around the excision boundaries covered by
  spherical grid is
  indicated by arrows.  The excision boundaries themselves
  are marked by circles drawn with dashed line.}
\end{figure}

In our spectral evolution code, the use of overlapping grids sometimes
leads to weak instabilities.  We find that these instabilities can be
cured by use of non-overlapping grids.  There are a number of choices
one has to make while designing such a grid. A basic assumption is
that at some distance from the center the geometry of the spacetime is
close to spherical symmetry.  Spherical shells are our most
efficient grid structure to represent such a region.  In the near zone
(around each singularity) we have an excision boundary of topology
$\Circle$ which suggests that, at least in the neighborhood of each
excision boundary, one can use spherical shells 
(see Fig.~\ref{fig:CutSphereShells}.)
Let $R_A$ and
$R_B$ be the outer radius of the region around the excision boundaries
that is described by spherical shells.
And let the coordinate centers of the
excision boundaries, as set by our initial data solver, be $(x_A,y_A,z_A)$ and 
$(x_B,y_B,z_B)$. Assume for the simplicity of the
discussion that $x_A>x_B$ and $|x_A|\leq|x_B|$.  We center the outer
shells at the origin of our coordinate system.  The inner radius $R_C$
of the outer spherical region is set to approximately three times 
the distance between the
centers of the excision spheres.
Next we need to fill in the space between the outer sphere
$\Sphere_C[(0,0,0),R_C]$ and the two inner spheres
$\Sphere_A[(x_A,y_A,z_A),R_A]$ and $\Sphere_B[(x_B,y_B,z_B),R_B]$.

In order to construct the actual subdomains filling up the
space between $S^3_A, S^3_B$ and $S^3_C$, we will make use
of $(\theta, \phi)$ coordinates aligned with the $x$ axis,
defined with respect to the centers of either $S^3_{EA}$ or $S^3_{EB}$
(these spheres will be defined below):
\begin{eqnarray}
\phi_A &=& \tan^{-1} ( z/y ), \\
\theta_A &=& \cos^{-1} \left(\frac{x-x_{EA}}{\sqrt{(x-x_{EA})^2+y^2+z^2}} \right)
\label{eq:CutXThetaPhi}
\end{eqnarray}
with similar definitions for $(\phi_B,\theta_B)$.

\begin{figure}
\centerline{\includegraphics[scale=0.4]{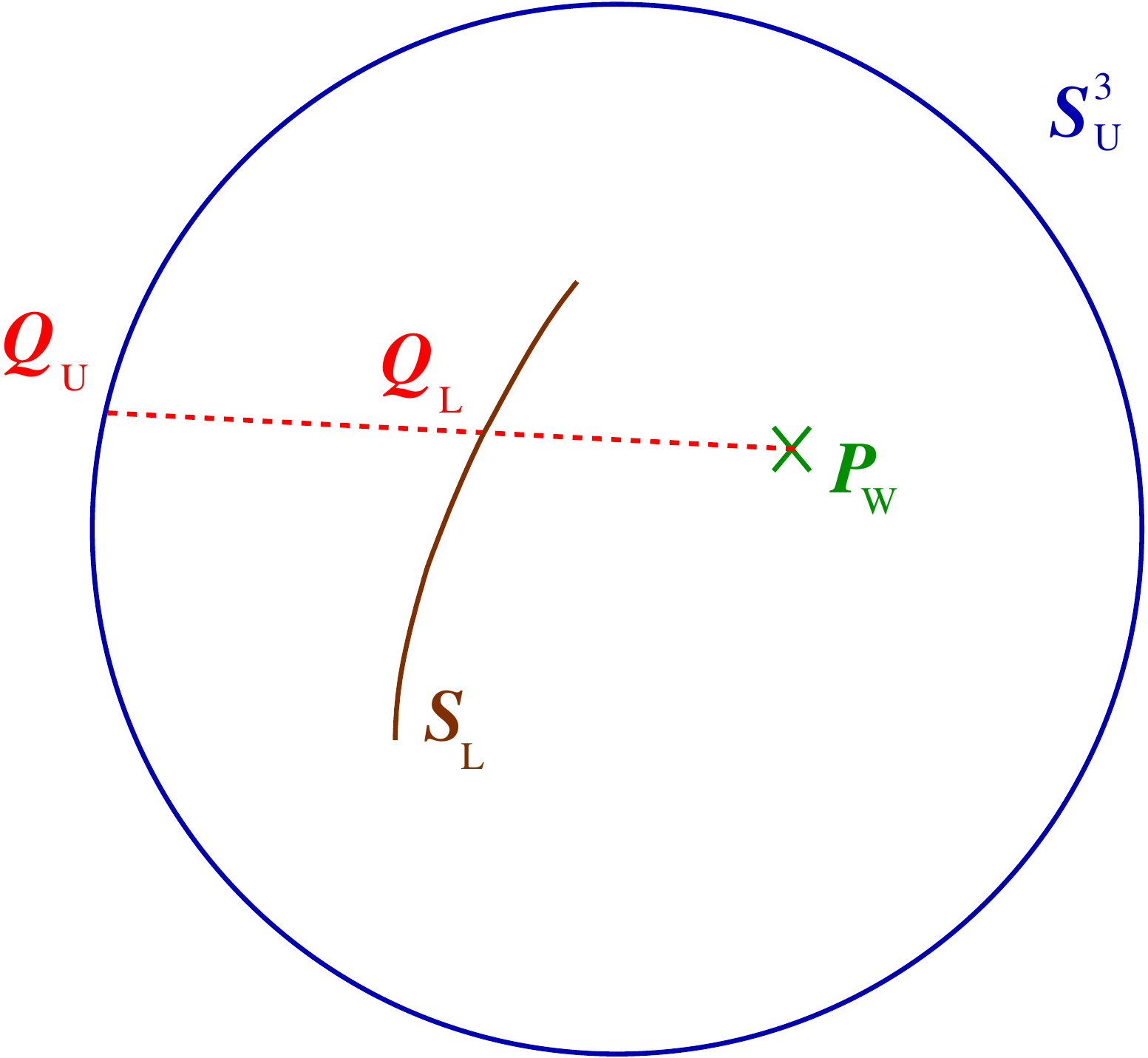}}
\caption{\label{fig:ProjectionMap} Schematic diagram
  of the projection map used to connect various surfaces
  with spheres. Given a a surface ${\cal S}_L$,
  a point $P_W$ and a sphere ${\cal S}^3_U$, the projection
  $Q_U$ of a point  $Q_L \in  {\cal S}_L$ is defined by
  the intersection of the line crossing $P_W, Q_L$
  and the sphere ${\cal S}^3_U$, such that $Q_L$ is 
  between $P_W$ and $Q_U$.
  }
\end{figure}

We next define a {\em projection map} used to connect
various surfaces with spheres (see Fig.~\ref{fig:ProjectionMap}).
Let ${\cal S}_L$
be a surface parametrized by $(\theta,\phi)$.
 Let ${\cal S}_U^3$ be a sphere, and
let $P_W$ be a point in the interior
of the sphere but not on the surface,
$P_W \not \in {\cal S}_L$.

For each point $Q_L(x_L^i) \in {\cal S}_L$
we construct a line connecting $Q_L$ and $P_W$.  This
will intersect the sphere in two points.
Let $Q_U(x_U^i)$ be the intersection point
that is on the same side of $P_W$ as $Q_L$.
Thus we have defined a rule that associates a
unique point $Q_U\in {\cal S}^3_U$ to
each point $Q_L \in  {\cal S}_L$.
We will
label the point $Q_U$ by the same parameters $(\theta,\phi)$
as the associated point $Q_L$.
The projection map
is defined as
\begin{eqnarray}
{\cal M}(P_W, {\cal S}_U^3)
&:=& \\ \nonumber
(\rho,\theta,\phi)
&\rightarrow& \frac{1-\rho}{2} x_L^i(\theta,\phi) + \frac{1+\rho}{2} x_U^i(\theta,\phi)
\end{eqnarray}
where we used $\rho$ as a radial parameter, with range $\rho\in[-1,1]$.  We have
\begin{eqnarray}
{\cal M}(-1,\theta,\phi) = x_L^i(\theta,\phi)
\\
{\cal M}(+1,\theta,\phi) = x_U^i(\theta,\phi)
\end{eqnarray}

We  associate one projection map with each of the
three spheres:
\begin{eqnarray}
{\cal M}_C &:=& {\cal M}\big((x_C,0,0), \Sphere_C\big) \\
{\cal M}_A &:=& {\cal M}\big(x_A^i, \Sphere_A\big) \\
{\cal M}_B &:=& {\cal M}\big(x_B^i, \Sphere_B\big)
\end{eqnarray}
where $x_C$ is defined in Eq.~(\ref{eq:CutSphereXC}).  As
  pointed out in Sec.~\ref{sec:ID}, $x^i_{A/B}$ are slightly offset
  from the $x$ axis along the $y$ direction.

\begin{figure}
\includegraphics[scale=0.367]{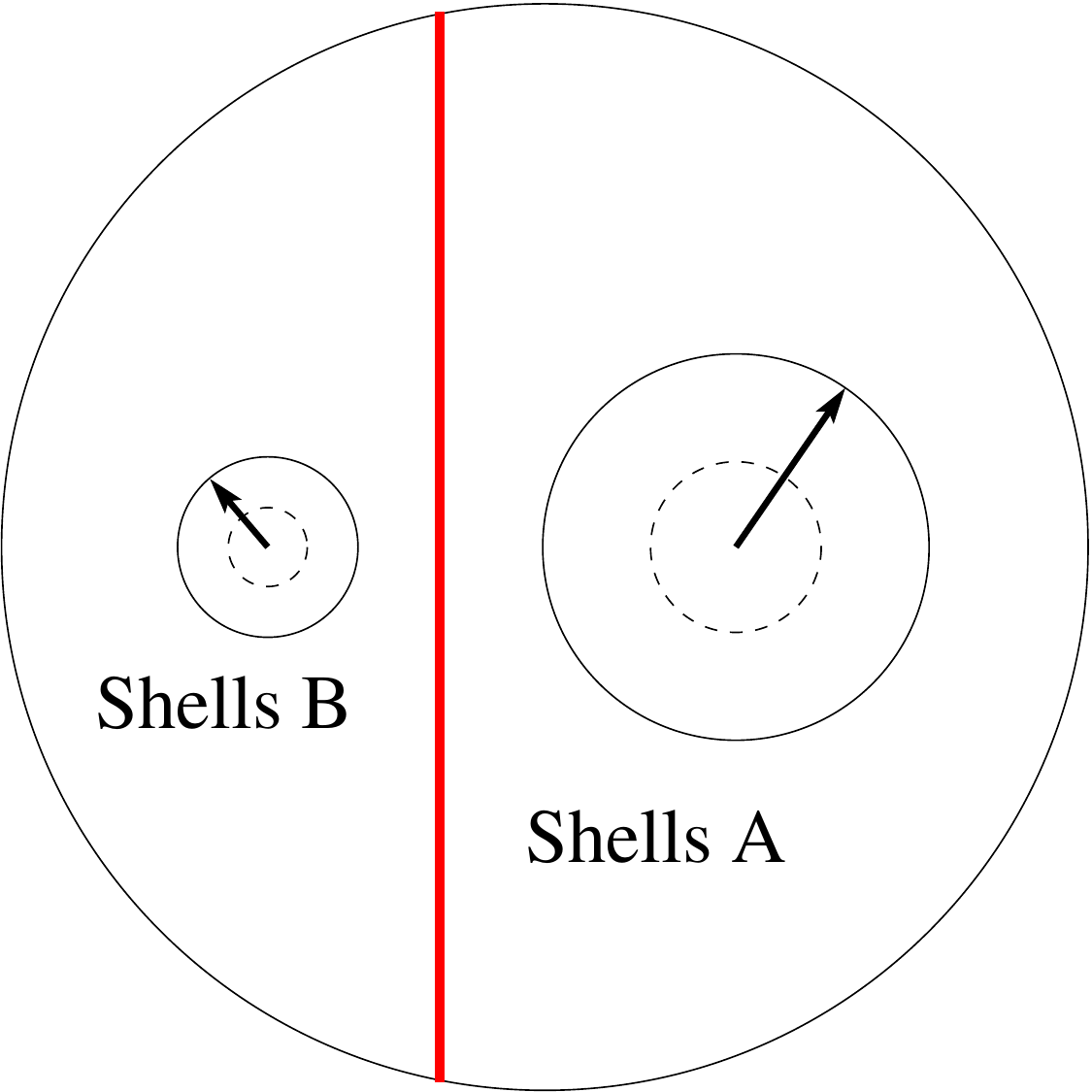}
\includegraphics[scale=0.185]{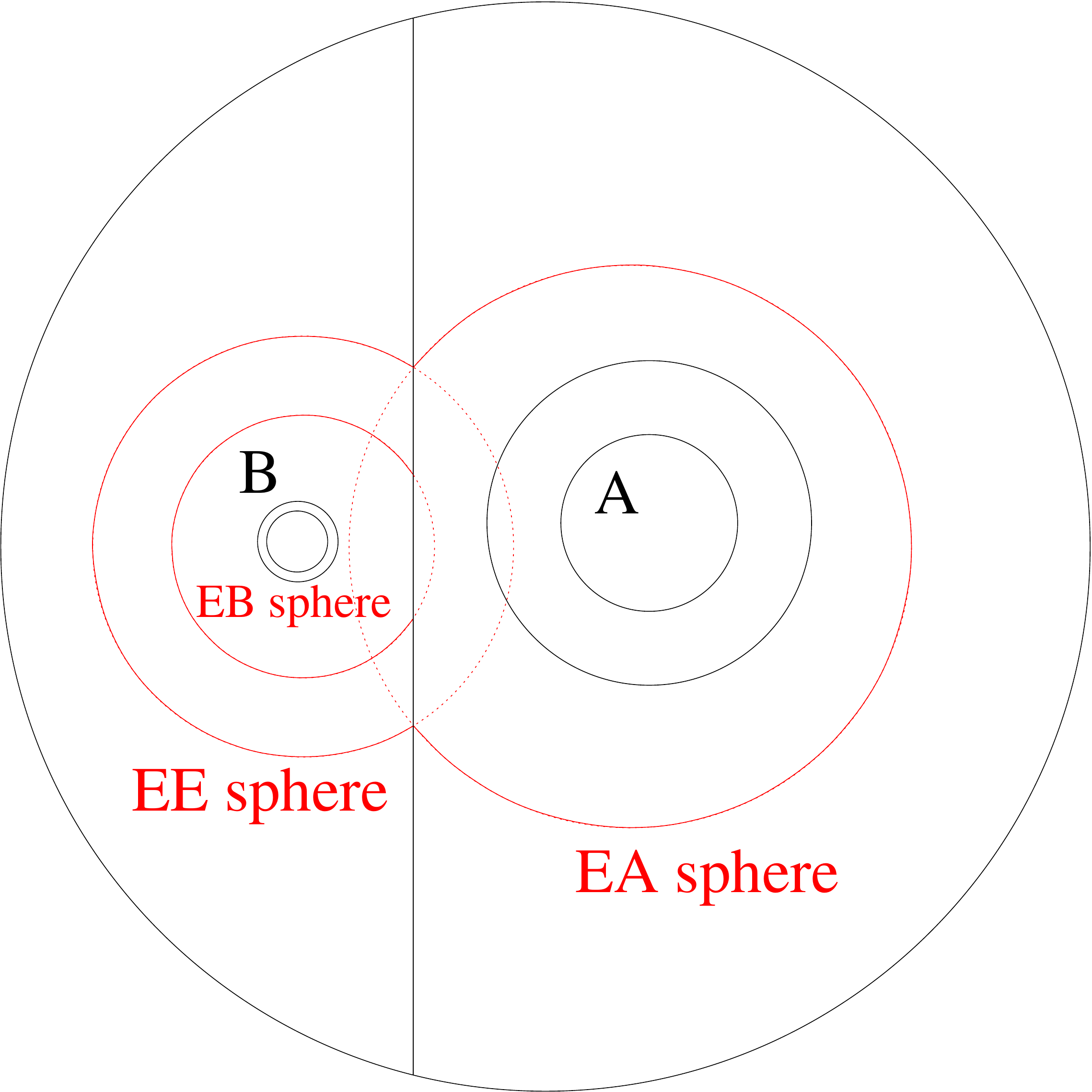}
\caption{\label{fig:CutSpherePlaneX}
  {\bf Left:} Schematic diagram indicating the $x=$const plane separating the
  regions around the two excision boundaries.  {\bf Right: }Schematic diagram of the spheres $\Sphere_{EA}, \Sphere_{EE}, \Sphere_{EB}.$}
\end{figure}

Next we divide the volume in the interior of $\Sphere_C$,
outside of $\Sphere_A$ and $\Sphere_B$ into wedges of various
shapes.
First we pick an $x=$const plane, $\PcutX$ (see Fig.~\ref{fig:CutSpherePlaneX}), that separates
the regions around the two excision boundaries,
using
\begin{eqnarray}
\label{eq:CutSphereXC}
x_C &=&  \eta (1-\xi) \, x_{A} + \eta\,  \xi x_{B}  ,\;\; \mbox{with}
\\
\xi &=& \max\left(\frac{1}{4},\frac{|x_{A}|}{|x_{A}|+|x_{B}|}\right)
\end{eqnarray}
Our preferred value for $\eta$ is $0.99$.

When $\xi \leq 1/3$ (corresponding to mass ratios $q\lesssim 2$)
we start by constructing a sphere 
$\Sphere_{\rm EA}\left[ (x_{\rm EA}, 0,0), R_{\rm EA}\right]$ with
\begin{eqnarray}
x_{\rm EA} &=& 0.9 \, \eta \, x_A \\
R_{\rm EA} &=& \sqrt{(x_{E\rm A}-x_C)^2+( \eta x_A - x_C )^2}.
\end{eqnarray}
The sphere $\Sphere_{\rm EA}$ intersects the plane $\PcutX$ in a circle
\begin{equation}
\Circle_{\rm ME} := \Sphere_{\rm EA} \cap \PcutX
\end{equation}
with radius 
\begin{equation}
r_{\rm ME}=|\eta x_A - x_C|.
\end{equation}
On the other side of $\PcutX$ we define
two concentric spheres (see Fig~\ref{fig:CutSpherePlaneX}):
$\Sphere_{\rm EB}\left[ (x_{\rm EB}, 0,0), R_{\rm EB}\right]$ and
$\Sphere_{\rm EE}\left[ (x_{\rm EB}, 0,0), R_{\rm EE}\right]$ 
 with
\begin{eqnarray}
x_{\rm EB} &=& \eta \, x_B \\
r_{\rm MB} &=& r_{\rm ME} \times
            \max \left( 0.4,
              \left| \frac{\eta\, x_B-x_C}{\eta\, x_A-x_C} \right |\right) \\
R_{\rm EE} &=& \sqrt{(x_{\rm EB}-x_C)^2+r_{\rm ME}^2} \\
R_{\rm EB} &=& \sqrt{(x_{\rm EB}-x_C)^2+r_{\rm MB}^2} .
\end{eqnarray}

\begin{figure}
\centerline{\includegraphics[scale=0.37]{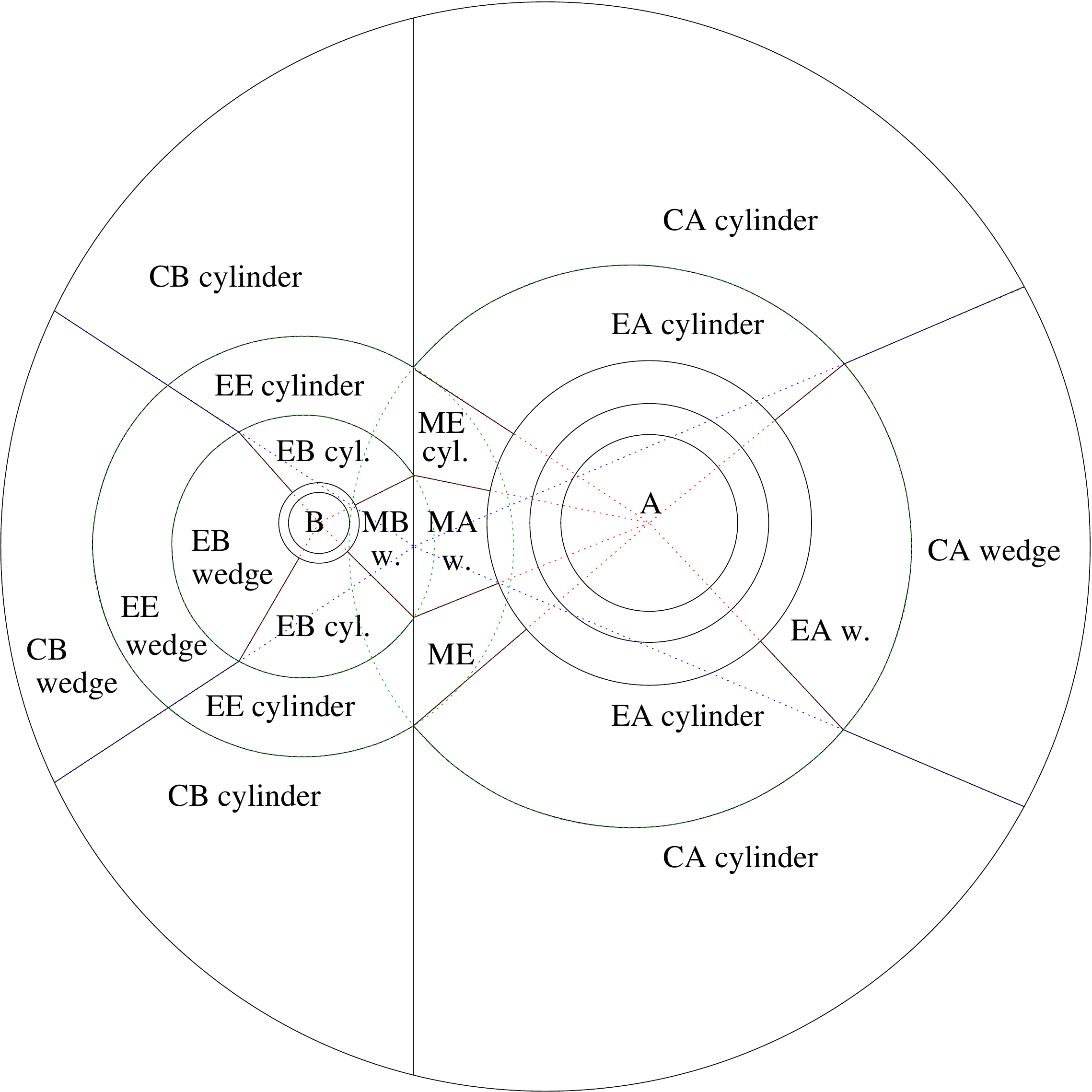}}
\caption{\label{fig:CutSphere} Schematic geometry of the touching
  grid geometry.  In a typical simulation we surround the shown
  grid geometry by about 20 further spherical shells on the outside,
  which are not shown in this diagram.}
\end{figure}

These choices imply that $\Sphere_{\rm EB}$ intersects $\PcutX$ in a circle 
with radius $r_{\rm MB}$
\begin{equation}
\Circle_{\rm MB}:= \Sphere_{\rm EB} \cap \PcutX.
\end{equation}
Next we define wedges/cylinders filling up the space between 
the three spherical surfaces.  (In our terminology
{\em wedges} have
topology $I^1\times B^2$, {\em cylinders} have topology 
$I^1 \times S^1 \times I^1$.)
\begin{itemize}
\item we connect the $x\geq x_C+(3/2) (x_{\rm EA}-x_C)$ portion of
      $\Sphere_{\rm EA}$ with $\Sphere_C$ using ${\cal M}_C$
      and call this the {\em CA wedge}
\item we connect the same portion of
      $\Sphere_{\rm EA}$ with $\Sphere_A$ using ${\cal M}_A$
      and call this the {\em EA wedge}
\item we connect the $x_C \leq x\leq x_C+(3/2) (x_{\rm EA}-x_C)$ portion of
      $\Sphere_{\rm EA}$ with $\Sphere_C$ using ${\cal M}_C$
      and call this the {\em CA cylinder}
\item we connect the same portion of
      $\Sphere_{\rm EA}$ with $\Sphere_A$ using ${\cal M}_A$
      and call this the {\em EA cylinder}
\item we connect the points $x^i\in \PcutX$ inside $\Circle_{\rm ME}$
      but outside $\Circle_{\rm MB}$  with $\Sphere_A$ using ${\cal M}_A$
      and call this the {\em ME cylinder}
\item we connect the points $x^i\in \PcutX$ inside $\Circle_{\rm MB}$
      with $\Sphere_A$ using ${\cal M}_A$
      and call this the {\em MA wedge}
\item we connect the same set of points
      with $\Sphere_B$ using ${\cal M}_B$
      and call this the {\em MB wedge}
\item we connect the $x\leq x_C-(3/2) |x_{\rm EB}-x_C|$ portion of
      $\Sphere_{\rm EB}$ with $\Sphere_C$ using ${\cal M}_C$.
      The portion inside $\Sphere_{\rm EE}$ is the {\it EE wedge},
      the portion between $\Sphere_{\rm EE}$ and $\Sphere_C$ is the 
      {\em CB wedge.}
\item we connect the same portion of
      $\Sphere_{\rm EB}$ with $\Sphere_B$ using ${\cal M}_B$
      and call this the {\em EB wedge}.
\item we connect the $x_C \geq x\geq x_C-(3/2) |x_{\rm EB}-x_C|$ portion of
      $\Sphere_{\rm EB}$ with $\Sphere_C$ using ${\cal M}_C$.
      The portion inside $\Sphere_{\rm EE}$ is the {\it EE cylinder},
      the portion between $\Sphere_{\rm EE}$ and $\Sphere_C$ is the 
      {\em CB cylinder.}
\item we connect the same portion of
      $\Sphere_{\rm EB}$ with $\Sphere_B$ using ${\cal M}_B$
      and call this the {\em EB cylinder}.
\end{itemize}

In the cases where $\xi > 1/3$ 
(corresponding to mass ratios $q\gtrsim 2$)
we use a slightly
simpler algorithm: we start by constructing
$\Sphere_{\rm EB}\left[ (x_{\rm EB}, 0,0), R_{\rm EB}\right]$ with
\begin{eqnarray}
x_{\rm EB} &=& \eta \, x_B \\
R_{\rm EB} &=& \sqrt{2} \times \left|x_{\rm EB} - x_C \right|.
\end{eqnarray}
The sphere $\Sphere_{\rm EB}$ intersects $\PcutX$ in a circle
\begin{equation}
\Circle_{\rm MB}:=\Sphere_{\rm EB} \cap \PcutX
\end{equation}
with radius
\begin{equation}
r_{\rm MB} =|\eta x_B - x_C|.
\end{equation}
On the other side of $\PcutX$ we define
\begin{eqnarray}
x_{\rm EA} &=& \eta \, x_A \\
R_{\rm EA} &=& \sqrt{(x_EA - x_C )^2 + r_{\rm MB}^2}.
\end{eqnarray}
Once again, $\Sphere_{\rm EA}\left[ (x_{EA}, 0,0), R_{\rm EA}\right]$ 
intersects $\PcutX$ in a circle
\begin{equation} 
\Circle_{\rm MB} := \Sphere_{\rm EA}\cap \PcutX.
\end{equation}

The definition of the various wedges and cylinders in this case
is similar to what is used for $\xi \leq 1/3$ with the exception
that there are no {\em EE} or {\em ME cylinders/wedges}, as 
$\Sphere_{\rm EA} \cap \PcutX = \Sphere_{\rm EB} \cap \PcutX =\Circle_{\rm MB}$.

See Fig.~(\ref{fig:CutSphere3D}) for a 3D snapshot of a grid
used for a run with mass ratio $2$.  This simulation uses the more complicated 
domain decomposition, although it is close to the
dividing line $\xi=1/3$ where we switch to the simpler domain decomposition.
As a last remark, in the runs described here, we have subdivided
each wedge (of topology $I^1\times B^2$) into five distorted
cubes.

%%%%%%%%%%%%%%%%%%%%%%%%%%%%%%%%%%%%%%%%%%%%%%%%%%%%%%%%%%%%%%%%%%%%%%%%%%%%%%%
%\section*{References}
%%%%%%%%%%%%%%%%%%%%%%%%%%%%%%%%%%%%%%%%%%%%%%%%%%%%%%%%%%%%%%%%%%%%%%%%%%%%%%%

\bibliography{References/References}

\end{document}